\documentclass[aps,prd,nofootinbib,twocolumn,superscriptaddress,showpacs,preprintnumbers]{revtex4-1}
\usepackage{epsfig,dcolumn}
\usepackage{graphicx}
\usepackage{color}

\usepackage{amsmath}
\usepackage{amsfonts}
\usepackage{amssymb}
\usepackage{graphicx}
\usepackage{bm} 
\usepackage[percent]{overpic}
 
\newcommand{\be}{\begin{equation}}
\newcommand{\ee}{\end{equation}}

\DeclareMathOperator{\re}{Re}
\DeclareMathOperator{\im}{Im}

\begin{document}

\title{Toward Complete Pion Nucleon Amplitudes}

\author{V.~Mathieu}
\email{mathieuv@indiana.edu}
\affiliation{Center for Exploration of Energy and Matter, Indiana University, Bloomington, IN 47403}
\affiliation{Physics Department, Indiana University, Bloomington, IN 47405, USA}
\author{I.~V.~Danilkin}
\affiliation{Thomas Jefferson National Accelerator Facility, 
Newport News, VA 23606, USA}
\author{C.~Fern\'andez-Ram\'{\i}rez}
\affiliation{Thomas Jefferson National Accelerator Facility, 
Newport News, VA 23606, USA}
\author{M.~R.~Pennington}
\affiliation{Thomas Jefferson National Accelerator Facility, 
Newport News, VA 23606, USA}
\author{D.~Schott}
\affiliation{Thomas Jefferson National Accelerator Facility, 
Newport News, VA 23606, USA}
\affiliation{Department of Physics, The George Washington University, Washington, DC 20052, USA}
\author{Adam~P.~Szczepaniak}
\affiliation{Center for Exploration of Energy and Matter, Indiana University, Bloomington, IN 47403}
\affiliation{Physics Department, Indiana University, Bloomington, IN 47405, USA}
\affiliation{Thomas Jefferson National Accelerator Facility, 
Newport News, VA 23606, USA}
\author{G. Fox}
\affiliation{School of Informatics and Computing, Indiana University, Bloomington, IN 47405, USA} 
 
\collaboration{Joint Physics Analysis Center}
\preprint{JLAB-THY-15-2056}

\begin{abstract}
We compare the low-energy partial wave analyses  $\pi N$ scattering with a high-energy data via finite energy sum rules. We construct a new set of amplitudes by matching the imaginary part from the low-energy analysis with the high-energy, Regge parametrization  and reconstruct the real parts using dispersion relations. 
\end{abstract}


\maketitle

\section{introduction}

Recent observations of several new hadron resonances,  including states that do not fit quark model expectations, demonstrate there is a significant discovery potential in the hadron spectrum~\cite{Klempt:2009pi}. On the theoretical side lattice gauge simulations have been evolving rapidly and simulations of hadron scattering will in the future provide first principle insights into the hadron spectrum and its QCD origins~\cite{Dudek:2012xn,Dudek:2014qha}. 
The common feature of data analysis and lattice simulations is that both require determination of reaction amplitudes.  Properties of known baryon resonances have been extracted from analysis of pseudoscalar-nucleon scattering and, more recently,  from single and double meson photoproduction~\cite{Anisovich:2011fc,Kamano:2013iva}. At present, the  properties of the nucleon and delta resonances with masses below $\sim 1.7$~GeV are quite well-determined~\cite{Beringer:1900zz} there is, however, a significant  number  of  high mass  resonances with questionable status and/or poorly known characteristics.  Since resonances appear as singularities of partial wave amplitudes in the complex energy and/or angular momentum planes, extraction of resonance parameters requires analytic continuation of reaction amplitudes outside the experimentally  accessible range of kinematical variables. This in turn implies that amplitudes should be  constrained as much as possible using principles of the analytic $S$-matrix \cite{Battaglieri:2014gca}. Specifically, amplitudes reconstructed from the low-energy partial wave analyses, that contain direct channel resonance dynamics, should smoothly connect with the high-energy region. The latter carry information about Regge poles and/or cuts exchanged in cross channels. The possibility that in the high-energy limit Regge poles dominate over  Regge cuts is particularly attractive given the factorization properties of the former. 

Practical implementation of matching between the low- and the high-energy domains explores analyticity of the reaction amplitude via dispersion relations. A summary of past work on Reggeized partial wave analysis can be found, for example in~\cite{Lovelace1,Lovelace:1970kn}. Dispersion relations can be used  in various ways. For example, the real part of an amplitude can be computed from the imaginary part and compared with alternative parameterizations {\it e.g.} Breit-Wigner, K-matrix or Chew-Mandelstam formulas. This approach was adopted by the SAID group in ~\cite{Arndt:1994bu,Arndt:1995bj,Arndt:2003if}.  Another option is, given low-energy amplitudes, to  use dispersion relations to extract parameters of Regge exchanges and compare them with those obtained from direct fits to the high-energy data. In this paper we explore both approaches. Specifically, we study dispersion relations in energy at fixed momentum transfer and moments of the amplitudes, {\it i.e.} integrals over energy. Dispersion relations applied to the moments lead to sum rules, so called finite energy sum rules, (FESR's)  that relate  the low- to the high-energy contributions to the amplitudes~\cite{Dolen:1967zz}.   FESR's provide stronger constrains than dispersion relations alone as they represent derivatives of the latter. By choosing appropriate moments, one can weight differently various domains of the low energy regions. 
  
We focus our analysis on $\pi N$ scattering as it is the building block for various analyses including multiple meson production. Currently the vast majority of  analyses use directly SAID elastic $\pi N$ amplitudes~ \cite{Workman:2012hx}. Unfortunately, fine details,  such as the Regge parametrization for the high energy region, are hard to find ~\cite{Arndt:1994bu,Arndt:1995bj,Arndt:2003if}. The purpose of this work is therefore two fold. Given that, at present, majority of amplitude analyses focus on the low-energy side we discuss in detail  the high-energy parametrization, fits, and the connection between the low-energy and the high-energy analyses. We emphasize application of dispersion relations and FESR's as a tool for constraining low-energy amplitudes  and ultimately for extraction of resonance parameters. As a result we can provide a set of amplitudes  valid in the high energy domain that can be used, via dispersion relations and FESR's to constrain phase shift analysis in the low energy region.\footnote{A website with on-line tools will be available~\cite{website}.}

The paper is organized as follows. The core of the approach  was developed in~\cite{Chew:1957zz, Bransden:1965zz} and in Sec.~\ref{sec:formalism} we summarize the relevant parts of the formalism. Description of amplitudes in the low-energy and in the high-energy parametrization are given in Sec.~\ref{sec:low} and Sec.~\ref{sec:high}, respectively. In Section~\ref{sec:low} we determine the contribution of the low-energy partial waves to the FESR's.  In the high-energy fits we determine the  $\rho$ Regge trajectory parameters from the $\pi N$ charge exchange data and use elastic $\pi N$  scattering to determine the Pomeron and the $f_2$ exchanges. 
We compare, in Sec.~\ref{sec:comp}, the contribution to FESR's from our high-energy fits with the contribution from the low-energy partial waves obtained by  SAID~\cite{Workman:2012hx}.  In Section~\ref{sec:new} we analyze the contribution from the  Regge poles to the sum rules and show how to reconstruct real parts of the $\pi N$ amplitudes using simultaneously the  low- and the high-energy data. Specifically, we interpolate the imaginary part between the two domains and reconstruct the real part using dispersion relations. We compare the reconstructed partial waves with the input amplitudes from SAID. We summarize our results and outline future prospects in Section \ref{sec:summary}.

\section{The Formalism}\label{sec:formalism}

\subsection{Kinematics} 

We use the standard parametrization of the $\pi N$ scattering amplitude, $T$ in terms of scalar functions $A$ and $B$~\cite{Chew:1957zz} 
\be
T_{\lambda_2,\lambda_4} =  \bar u(p_4,\lambda_4) \left[ A_{ba}^{ji}  
+ \frac{1}{2} (p_1\!\!\!\!\! /\ +p_3\!\!\!\!\! / \ ) B_{ba}^{ji} \right] u(\lambda_2,p_2).  \label{0} 
\ee
The scalar amplitudes are functions of the standard Mandelstam variables, $s,t,u$, related by $s+t+u= 2M^2 + 2\mu^2$, where $M$ and $\mu$ refer to the nucleon and pion mass, respectively. In the $s$-channel, which corresponds to $\pi N \to \pi N$, $s=W^2$ is the square of the total energy in the center of mass frame and the other two variables, $t$ and $u$ are related to the scattering angle in this frame. Except when explicitly stated all quantities are given  in units of GeV. The $t$-channel corresponds to the reaction $\pi \pi \to N\bar N$. In Eq.~(\ref{0}) the indices $b,a$ and $i,j$ label the pion and the nucleon isospin, respectively.  The $A$ and $B$ amplitudes can be decomposed in terms of amplitudes with well defined total isospin in either $t$ or $s$ channel. In terms of the $t$-channel isospin amplitudes, denoted by $A^{(+)}$ for isospin-0    and $A^{(-)}$ for isospin-1, the amplitudes in Eq.(\ref{0}) are given by 
\be
A_{ba}^{ji} = \delta_{ba}\delta_{ji} A^{(+)} + i\epsilon_{bac}\left(\tau^c\right)_{ji} A^{(-)},
\ee
and similarly for the $B$ amplitude.  The relations between the $t$-channel and the $s$-channel, $\pi N \to \pi N$,  isospin-1/2 and isospin-3/2 amplitudes are,
\begin{align}
A^{(\frac{1}{2})} &= A^{(+)} + 2A^{(-)}, & A^{(\frac{3}{2})} &= A^{(+)} - A^{(-)}.
\end{align}
In the following, however,  we will be primarily working with the $t-$channel isospin amplitudes.
Partial wave expansion in the $s$-channel, which will be used below to parametrize the $A$ and $B$ amplitudes in the 
 nucleon resonance region, is written for the so-called reduced helicity amplitudes,  $f_1$ and $f_2$, which are related to $A,B$ by
\begin{subequations} \label{eq:AB2f}
\begin{align}
\frac{1}{4\pi} A^{(\pm)} & = \frac{W+M}{E+M} f^{(\pm)}_1 - \frac{W-M}{E-M} f^{(\pm)}_2, \\
\frac{1}{4\pi} B^{(\pm)} & = \frac{1}{E+M} f^{(\pm)}_1 + \frac{1}{E-M} f^{(\pm)}_2. \label{f12} 
\end{align}
\end{subequations}
Here $E= (s+M^2-\mu^2)/2W$ denotes the nucleon energy in the $s$-channel center of mass frame, 
The partial wave expansion  is given by \cite{Jacob:1959at}
\begin{align}\nonumber
f^{(\pm)}_1(s,t) &= \frac{1}{q}\sum_{\ell=0}^\infty f^{(\pm)}_{\ell+}(s) P_{\ell+1}^\prime(z) - \frac{1}{q}\sum_{\ell=2}^\infty  f^{(\pm)}_{\ell-}(s) P_{\ell-1}^\prime(z)\\ \label{eq:PW}
f^{(\pm)}_2(s,t) &= \frac{1}{q}\sum_{\ell=1}^\infty\left[ f^{(\pm)}_{\ell-}(s)  - f^{(\pm)}_{\ell+}(s)\right] P_{\ell}^\prime(z) 
\end{align}
with $f_{\ell\pm}$ being the partial wave amplitudes with parity $(-1)^{\ell+1}$ and total angular momentum $J=\ell\pm1/2$. 
Here, $z=\cos\theta_s=1+t/2q^2$  denotes cosine of the $s$-channel scattering angle and 
$q=\sqrt{E^2-M^2}$ is the relative momentum in the $s$-channel center-of-mass frame.  In this frame,  the helicity amplitudes are given by 
\begin{subequations} \label{eq:SCHA}
\begin{align}
T^{s(\pm)}_{++} &= 8\pi W  \left(\frac{1+z}{2}\right)^{\frac{1}{2}} \left(f^{(\pm)}_1+f^{(\pm)}_2\right),\\ \label{eq:SCHAhighB}
T^{s(\pm)}_{+-} & = 8\pi W  \left(\frac{1-z}{2}\right)^{\frac{1}{2}}  \left(f^{(\pm)}_1-f^{(\pm)}_2\right), 
\end{align}
\end{subequations}
where the lower-script $\pm$ stands for the nucleon helicity $\pm1/2$.  

At high energies and small angles the reaction is dominated by leading Regge singularities in the $t$-channel which are given in terms of the $t$-channel helicity amplitudes, {\it i.e.}
\begin{subequations} \label{eq:TCHA}
\begin{align}
T^{t(\pm)}_{++} &= -\sqrt{t-4M^2} A'^{(\pm)},\\
T^{t(\pm)}_{+-} & =  \frac{1}{2}\sqrt{t(t-4\mu^2)} \sin\theta_t B^{(\pm)}, 
\end{align}
\end{subequations}
where 
\begin{align}
A' \equiv A + \frac{M(s-u)}{4M^2-t} B,
\end{align}
and the scattering angle in the $t$-channel, $\theta_t$ satisfies,  
\begin{align}
\sin\theta_t = \frac{1}{2} \sqrt{\frac{su - (M^2-\mu^2)^2}{(t/4-\mu^2)(t/4-M^2)} }.
\end{align}
The amplitudes are normalized in a way that the total cross section, differential cross section and polarization asymmetry, are given by 
\begin{subequations}\label{eq:obs}
\begin{align} \nonumber
\sigma_{\text{tot}}&=  \frac{1}{2 qW}\ [T^s_{++} + T^s_{+-}]|_{t=0}, \\\label{eq:totsig}
& = \frac{\text{Im} A'(s,t=0)}{p_\text{lab}},\\ \nonumber
\frac{d\sigma}{dt}&=  \frac{\pi}{q^2} \left(\frac{1}{8\pi W}\right)^2\left(|T^s_{++}|^2 + |T^s_{+-}|^2\right),  \\ \nonumber
& = \frac{1}{\pi s} \left(\frac{M}{4q}\right)^2 \bigg[ \left(1- \frac{t}{4M^2}\right) |A'|^2 \\
& \qquad \quad - \frac{t}{4M^2} \left(\frac{st + 4M^2 p^2_\text{lab} }{4M^2-t} \right) |B|^2\bigg], \\
\nonumber
P &= \frac{2\,\text{Im} \ T^s_{++}T^{s*}_{+-}}{|T^s_{++}|^2 + |T^s_{+-}|^2} \\
&= -\frac{\sin\theta_s}{16\pi W} \frac{\text{Im}\left(A' B^*\right)}{d\sigma/dt}, 
\label{eq:obsc}
\end{align}
\end{subequations}
 with $p_\text{lab}$ being the initial pion momentum in the nucleon rest frame (the lab frame). 
The $A$ and $B$ amplitudes for the charge exchange reaction,  $\pi^-p\to\pi^0 n$ are related to the $t$-channel isospin amplitudes by, 
\begin{align}
A' &= -\sqrt{2}\ A'^{(-)}, & B &= -\sqrt{2}\ B^{(-)},
\end{align}
and elastic scattering,  $\pi^{\pm} p$, 
\begin{align}
A' &=A'^{(+)} \mp A'^{(-)}, & B &=B^{(+)} \mp B^{(-)}.
\end{align}

\subsection{Finite Energy Sum Rules} 
The invariant amplitudes $A$ and $B$ are free from kinematical singularities. The only singularities are those demanded by unitarity, which at fixed-$t$ are the $s-$ and the $u-$channel thresholds,  and the nucleon pole. This leads to the dispersion relations which we write for amplitudes with fixed 
 $t$-channel isospin~\cite{Chew:1957zz} 
\begin{subequations}\label{eq:FTDRAB}
\begin{align}
A^{(\pm)}(\nu,t) &= \frac{1}{\pi}\int_{\nu_0}^\infty \im A^{(\pm)}(\nu',t) \left(\frac{d\nu'}{\nu'-\nu}\pm \frac{d\nu'}{\nu'+\nu}\right), \\ \nonumber
B^{(\pm)}(\nu,t) &= \frac{g^2_r}{2M} \left(\frac{1}{\nu_M-\nu}\mp \frac{1}{\nu_M+\nu}\right)\\ 
& + \frac{1}{\pi}\int_{\nu_0}^\infty \im B^{(\pm)}(\nu',t) \left(\frac{d\nu'}{\nu'-\nu}\mp \frac{d\nu'}{\nu'+\nu}\right), 
\end{align}
\end{subequations}
The variable $\nu$ defined by 
\begin{align}\label{nu} 
\nu = \frac{s-u}{4M} = E_{\text{lab}} + \frac{t}{4M}\ge \mu +\frac{t}{4M}=\nu_0 
\end{align}
 is introduced to account for the $s-u$ crossing symmetry. In Eq.~(\ref{nu}), $E_\text{lab}$ is the pion energy in the nucleon rest frame  and $\nu_0 = \nu(E_\text{lab}=\mu)$ corresponds to the value at the $\pi N$ threshold. The contribution from the nucleon pole corresponds to $\nu_M = (t-2\mu^2)/4M$. The residue of the nucleon pole is proportional to the renormalized $\pi NN$  coupling constant, $g^2_r \approx  56\pi $\footnote{We use the value $g^2_r = 56\pi $ in our numerical evaluation}. $A^{(+)}$ and $B^{(-)}$ $\left(A^{(-)} \text{ and }B^{(+)}\right)$ are even (odd) under crossing. 
One often considers dispersion relations for the amplitudes $A'^{(\pm)}$ and $\nu B^{(\pm)}$. They are proportional to $t$-channel helicity amplitudes and thus have the asymptotic limit as $s\to \infty$ fixed by the leading Regge singularity of the $t$-channel partial waves.  
They correspond to amplitudes with  $t$-channel helicity non-flip ($A'$) and flip 
($\nu B$), respectively and with $t$-channel isospin $0$ (superscript $-$)  and $1$ (superscript $+$), respectively. 

In what follows we  summarize the derivation of the finite energy sum rules. The derivation applies to  $A'^{(\pm)}$ and  $\nu B^{(\pm)}$, with $F$ standing for either $A$ or $\nu B$, 
\begin{eqnarray}\label{eq:FTDR}
F^{\pm}(\nu,t) &= &  G_M \left(\frac{1}{\nu_M-\nu}\pm \frac{1}{\nu_M+\nu}\right) \nonumber \\
&  +  & \frac{1}{\pi}\int_{\nu_0}^\infty d\nu' \im F^{\pm}(\nu',t) \left(\frac{1}{\nu'-\nu}\pm \frac{1}{\nu'+\nu}\right). \nonumber \\
\end{eqnarray}
The nucleon pole lies outside the range of integration and is given  by the first term on the right hand side of Eq.~(\ref{eq:FTDR}). $G_M$ is the residue at the nucleon pole. Its value for specific amplitudes is
\begin{subequations}
\begin{align}
G_M &= \frac{\nu}{1-t/(4M^2)} \frac{g_r^2}{2M} & \text{for } F&=A', \\
G_M &= \nu\frac{g_r^2}{2M} & \text{for } F&=\nu B 
\end{align}
\end{subequations}

In the following we focus on the dispersive part. At fixed-$t$ we approximate the large-$\nu$ behavior of the amplitudes by $t$-channel Regge poles. Regge-pole contribution has the  form of 
\be\label{eq:Regge1}
R^{\tau}(\nu,t) = -\beta(t) \frac{e^{-i\pi\alpha} + \tau }{\sin\pi\alpha} \nu^\alpha,
\ee 
where $\tau = \pm 1$ is the signature.  In the physical region of the $t$-channel, amplitudes with positive (negative) signature correspond to exchanges of spin-even (odd) resonances, {\it e.g.} $\rho$ exchange has odd signature and the Pomeron and the $f_2$ have positive signatures. In Eq.~(\ref{eq:Regge1}), $\alpha = \alpha(t)$ is the  Regge pole trajectory and $\beta = \beta(t)$ is the residue. In the $s$-channel physical region, both are smooth functions of $t$.  In the derivation of the FESR it is assumed that $R^\pm $ is a good approximation to $F^\pm$ at high energies, {\it i.e.} $\nu \ge \Lambda$. The value of $\Lambda$ is to be chosen by comparing with the data. The function $R^\pm$ can be represented through a dispersive integral,  
\be\label{eq:FTDRReg}
R^{\pm}(\nu,t) =\frac{1}{\pi}\int_0^\infty d\nu' \im R^{\pm}(\nu',t) \left(\frac{1}{\nu'-\nu} \pm \frac{1}{\nu'+\nu} \right),
\ee
with $\im R^\pm(\nu) =  \beta(t) \nu^\alpha$.  Combining  Eq.~\eqref{eq:FTDRReg} with  Eq.~\eqref{eq:FTDR} 
 and approximating, for $\nu \ge \Lambda$, $\im F^\pm$  by the Regge amplitude  $\im R^\pm$ 
  one finds that for $\nu \ge \Lambda$ 
\begin{align} 
F^{\pm}(\nu,t) &=  R^\pm(\nu,t) - \sum_{k=0}^\infty \frac{1 \mp (-1)^k}{\nu^{k+1}} Q^{\pm}_{k}(\Lambda,t),
\end{align} 
where 
\begin{align*}
\pi Q^{\pm}_k(\Lambda,t) &\equiv  \pi G_B \nu_B^k \\
+ &  \int_{\nu_0}^\Lambda \im F^{\pm} (\nu,t)\nu^{k}d\nu
  -  \int_{0}^\Lambda \im R^{\pm} (\nu,t) \nu^{k}d\nu
\end{align*}
Finally,  equating $\re F^\pm$ with $\re R^\pm$ for $\nu \ge \Lambda$ leads to  the condition $Q^\pm_k(\Lambda,t) = 0$, and therefore,  
\begin{align} \nonumber 
\pi G_B \left(\frac{\nu_B}{\Lambda}\right)^k + \frac{1}{\Lambda^{k}}\int_{\nu_0}^\Lambda \im F^{\pm} (\nu,t)  \nu^{k}d\nu &= \frac{\beta(t)\Lambda^{\alpha+1}}{\alpha+k+1} \\  \label{eq:SR}
&\equiv S^\pm_k(\Lambda,t),
\end{align}
with odd (even) $k$ entering the sum rule for $F^+$ and $F^-$, respectively.  This sum rule  relates
  integrals over the imaginary part of the amplitudes $F^\pm$ taken over the low energy region, $\nu < \Lambda$ on left hand side to the parameters of the Regge singularities in the cross-channel on the right hand side.

\section{Application of FESR} 
In this section we evaluate the sum rules. To evaluate the left hand side ({\it l.h.s.}) we use various low energy ($\nu \le \Lambda)$ parametrizations  and for the right hand side ({\it r.h.s.}) we use a Regge-pole fit to the high-energy data. 
   
\subsection{Low Energy Parametrization}\label{sec:low}
\begin{figure*}[htb]\begin{center}
	\begin{overpic}[width=0.45\textwidth]{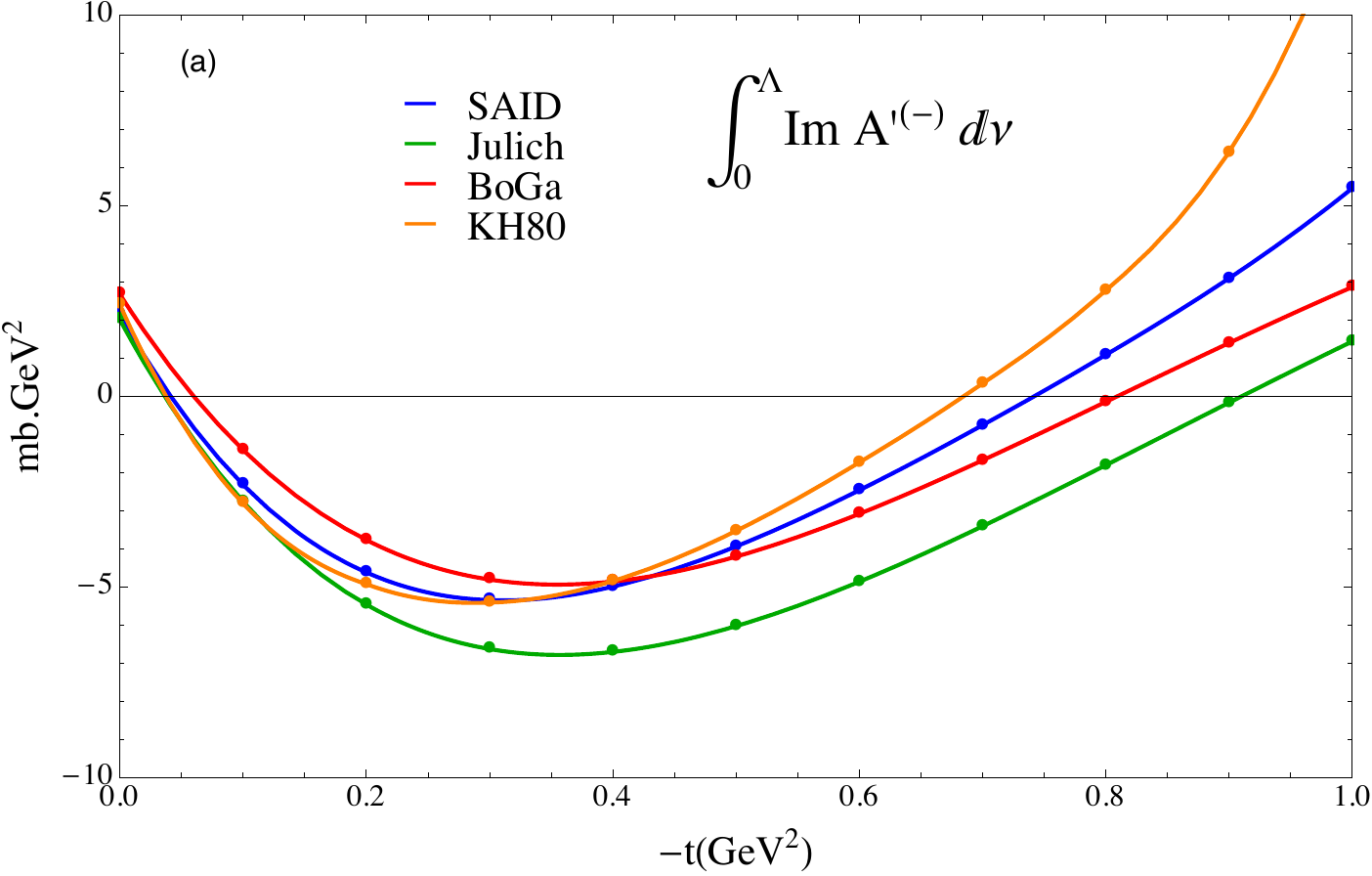}
 		\put (10,10) {\large$(a)$}
	\end{overpic}
	\begin{overpic}[width=0.45\textwidth]{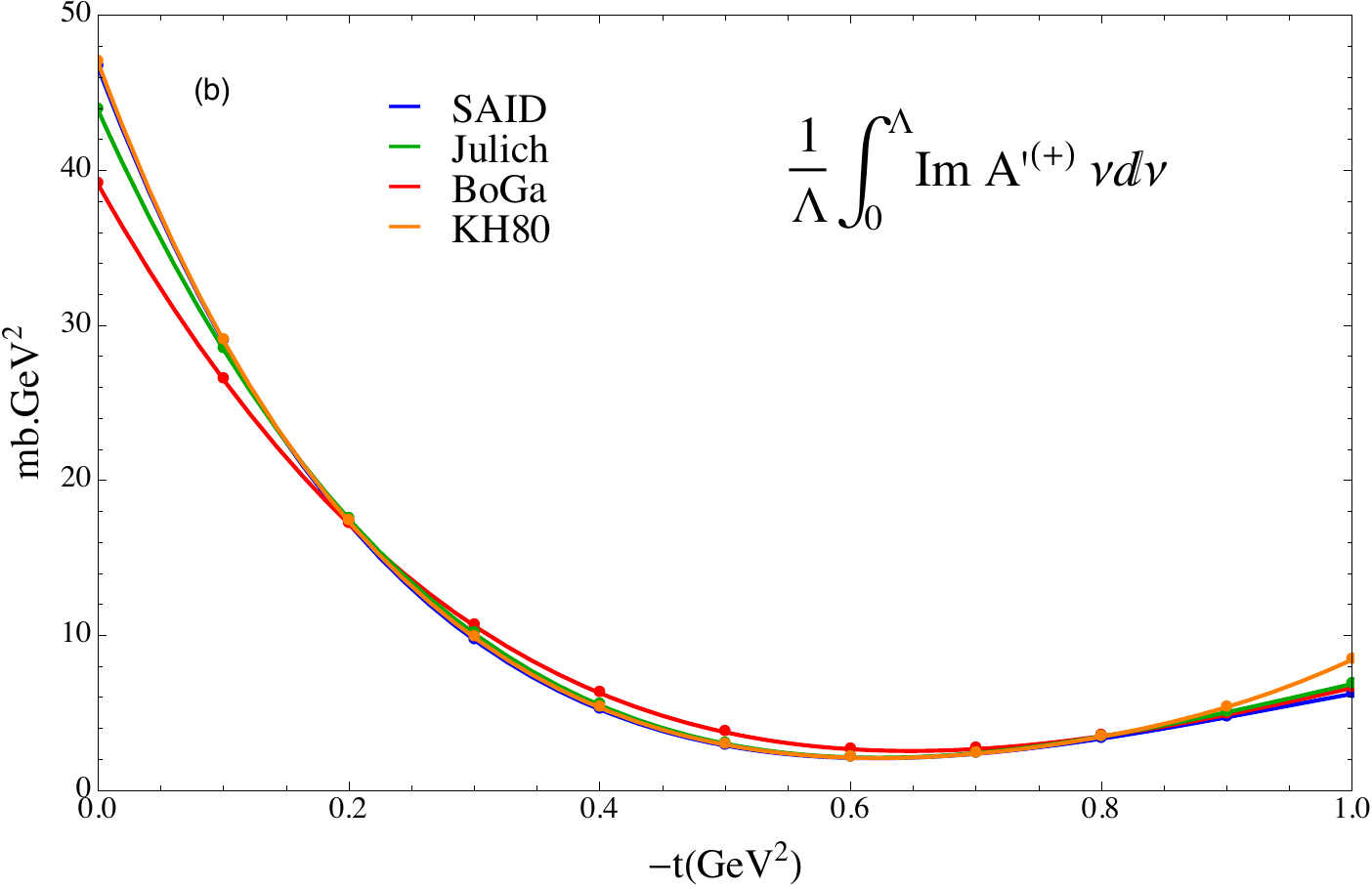}
 		\put (10,10) {\large$(b)$}
	\end{overpic}
	\begin{overpic}[width=0.45\textwidth]{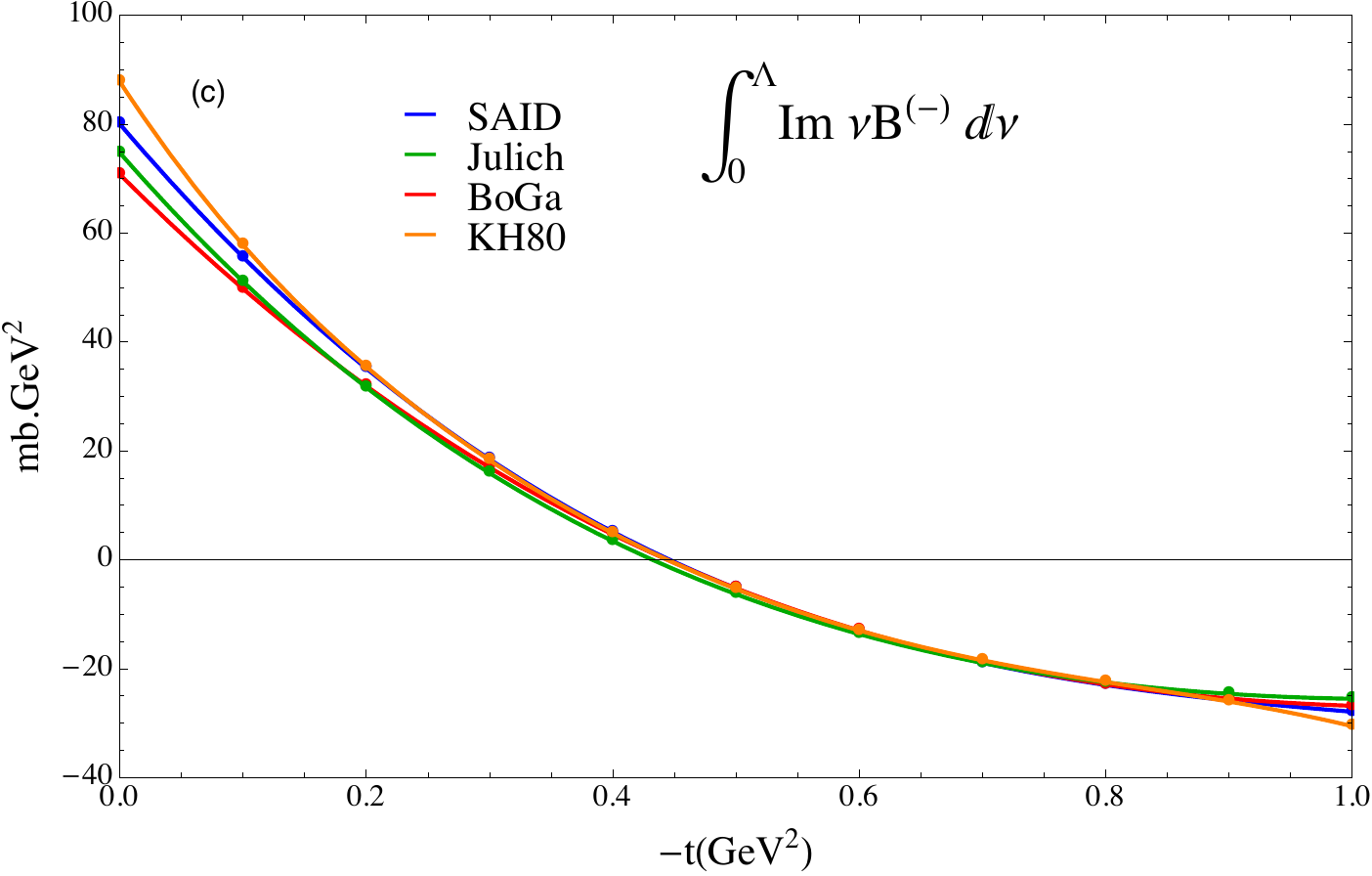}
 		\put (10,10) {\large$(c)$}
	\end{overpic}
	\begin{overpic}[width=0.45\textwidth]{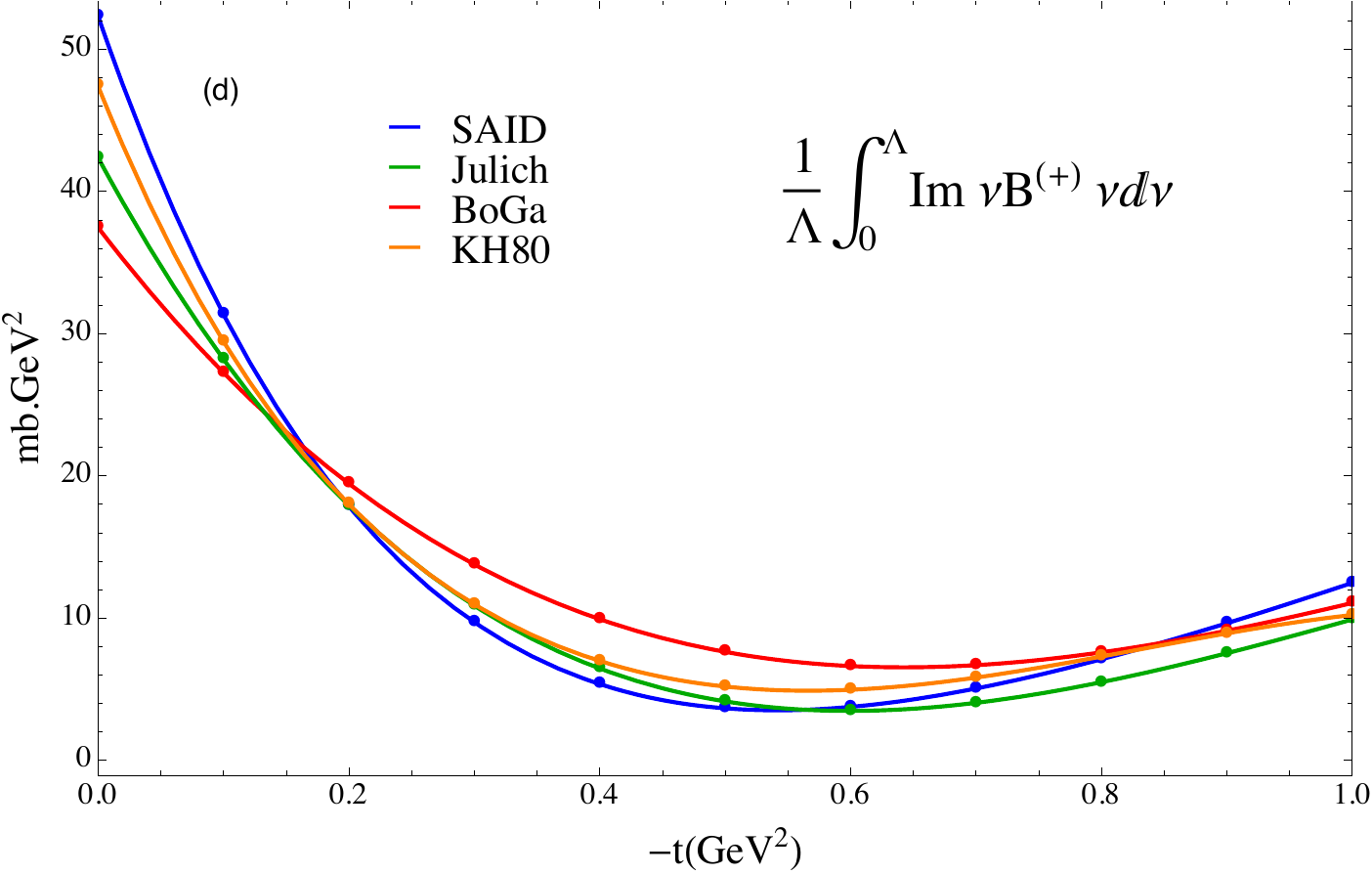}
 		\put (10,10) {\large$(d)$}
	\end{overpic}
\caption{\label{fig:FESRSolt} (color online) Left hand side of the sum rule, in Eq.\eqref{eq:SR},  computed with $k=0$ for crossing odd amplitudes $A'^{(-)}$ and $\nu B^{(-)}$ and with $k=1$ for crossing even amplitudes $A'^{(+)}$ and $\nu B^{(+)}$ using the four low-energy parametrizations discussed in the text and  $E^\text{max}_\text{lab}=2$ GeV. Fig (a) amplitude $A'^{(-)}$. Fig (b) amplitude $\nu B^{(-)}$. Fig (c) amplitude $A'^{(+)}$. Fig (d) amplitude $\nu B^{(+)}$.}
\end{center}
\end{figure*}

The left hand side of the FESR in Eq.~\eqref{eq:SR}  can be evaluated using the low-energy partial wave expansion. The invariant amplitudes are computed from partial waves using Eqs.~\eqref{eq:PW}. In the following we compare amplitudes obtained by SAID, \cite{Workman:2012hx}  (specifically the WI80 solution), Bonn-Gatchina (BoGa) \cite{Anisovich:2012ct}, the Julich model \cite{Ronchen:2012eg} (more precisely, the fit A), and the Karlsruhe-Helsinki  (KH80 solution)  \cite{KH80}. The Bonn-Gatchina and Julich analyses are coupled channel analyses. Their parametrization of the $\pi N\to \pi N$ channel is determined by fitting the SAID solution. The  BoGa, Julich and KH80 amplitudes are binned in $W$, starting from $W=1080$ MeV with 5 MeV bins (BoGa, Julich) and 10 MeV bins (KH80). The SAID amplitudes are binned in $E_\text{lab}$, starting from $E_\text{lab}=10$ MeV with 10 MeV bins. We use cubic spline interpolation between bins in numerical evaluation of the amplitudes. 
  
The Bonn-Gatchina and Julich analyses include partial waves ({\it cf.} Eq.~(\ref{eq:PW})) with angular momentum up to $\ell=4$ and $\ell=5$, respectively, while the SAID and KH80 include waves up to $\ell=7$. For $t$ outside the physical region of the $s$-channel, invariant amplitudes are obtained by analytical continuation. Continuation outside the $s$-channel physical region ($|z_s| = |\cos\theta_s| > 1$) based on a truncated set of partial waves in general produces unphysical results. At fixed $s$ (or $\nu$) invariant amplitudes obtained this way become less reliable as the magnitude of $z_s$ or $t$ increase.  Using the four sets of partial wave amplitudes  we typically find that, as long as $\ell_{max} \le 7$, for $s>1.2$ GeV$^2$ the contribution to the FESR that originates from integration over $\nu$ in the unphysical region is stable as long as  $|t|$ is smaller than $1$ GeV$^2$. Therefore, in computation of the left hand side of the FESR we restrict the range of $t$ to  $-1 \le t \le 0$ GeV$^2$.  Alternative methods for extending the range of applicability of the truncated partial wave sum were discussed, for example, in \cite{FerroFontan:1972in}. We do not follow them here since the simple truncation gives a stable result  when extrapolated to restricted range of $t$. 

From Eq.~\eqref{nu} it follows that the cutoff $\Lambda$ which enters the expression for $S_k(\Lambda,t)$, in Eq.~\eqref{eq:SR} depends on the  beam energy $E_\text{lab}$ and $t$, $\Lambda=E^\text{max}_\text{lab}+t/4M$. All four partial wave solutions are constrained by data up to (at least) $E_\text{lab}=2.1$ GeV. In the study of the FESR we therefore use $E^\text{max}_\text{lab}=2$~GeV when determining the cutoff. 

The left hand side of the sum rule, Eq.~\eqref{eq:SR} is a function of $t$ determined by integrals over the low energy partial waves. The sum rule relates this $t$-dependence to that of the Regge pole parameters appearing on the right hand side. For example, vanishing of $S_k(\Lambda,t)$  at a particular value of $t$  on the left hand side  would imply a zero in the residue $\beta(t)$, if the right hand side were dominated by a single Regge pole. In general, however, the right hand side receives contributions from more than one Regge-pole and    matching $t$ dependencies of the two sides of the sum rule is not so simple. The Regge pole parametrization of the right hand side will be discussed in the following section. Here we comment on the features of the  $t$-dependence observed for the left hand side of the sum rule.

\begin{figure*}[htb]\begin{center}
	\begin{overpic}[width=0.45\textwidth]{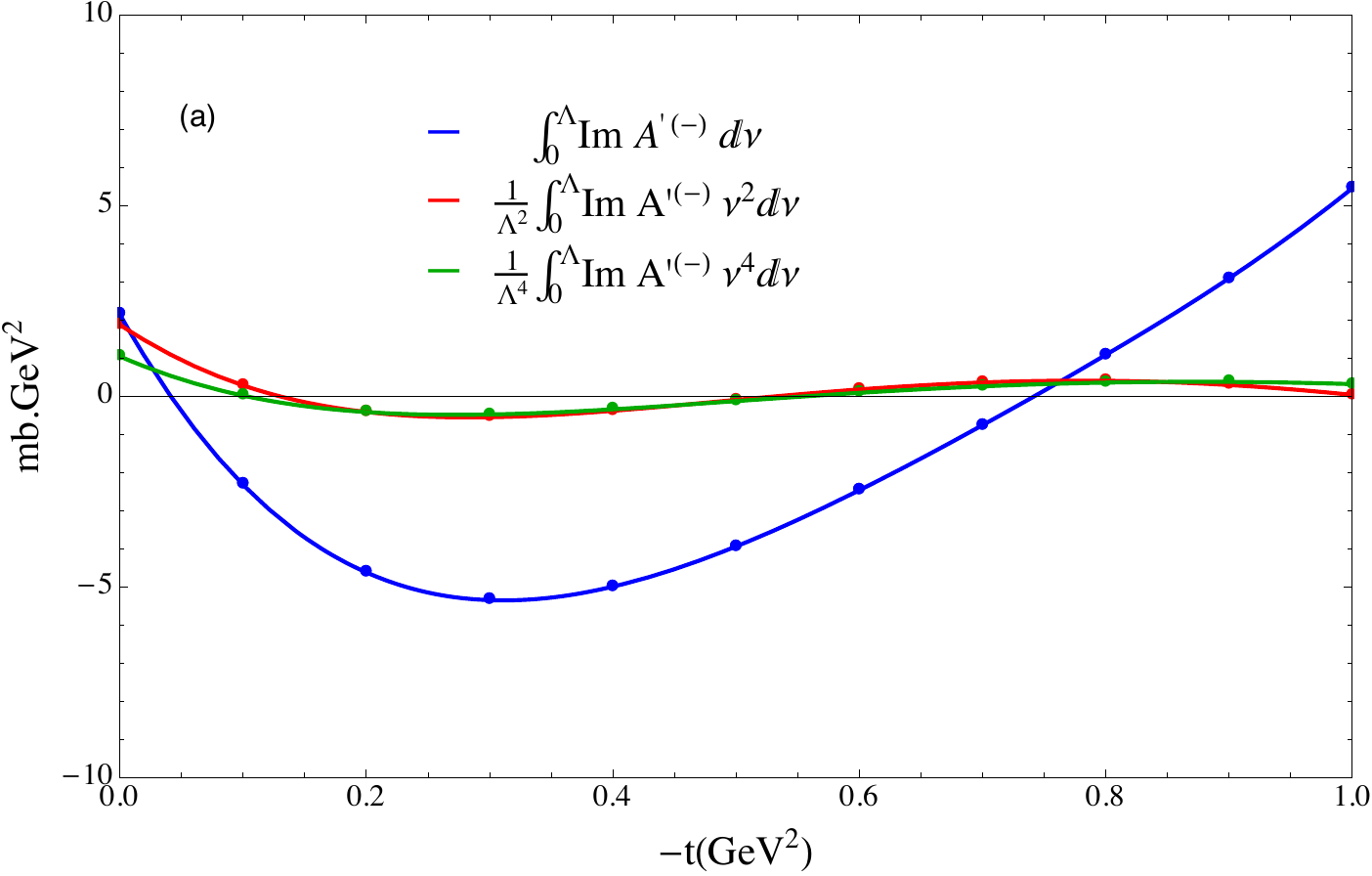}
 		\put (10,10) {\large$(a)$}
	\end{overpic}
	\begin{overpic}[width=0.45\textwidth]{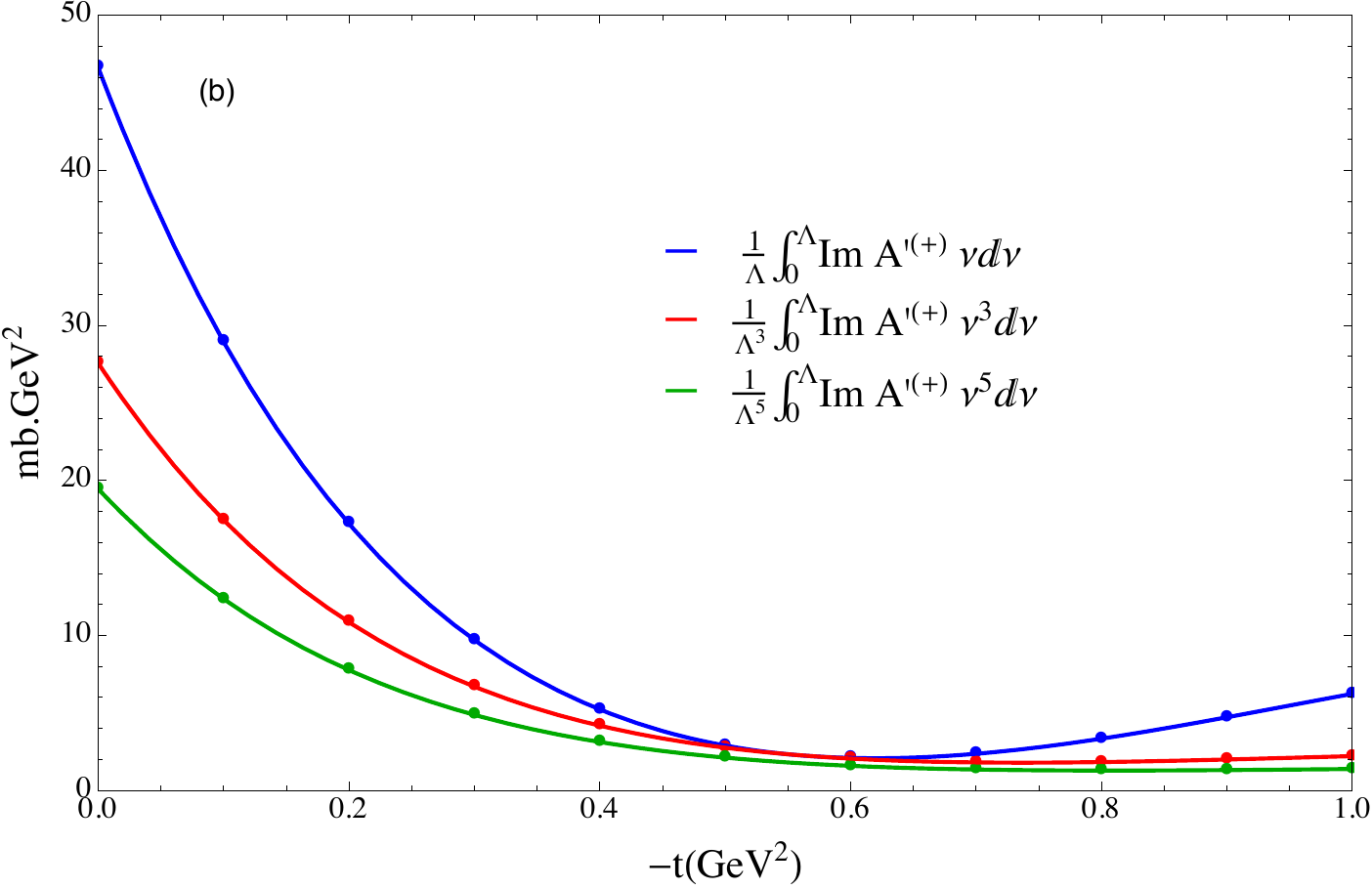}
 		\put (10,10) {\large$(b)$}
	\end{overpic}
	\begin{overpic}[width=0.45\textwidth]{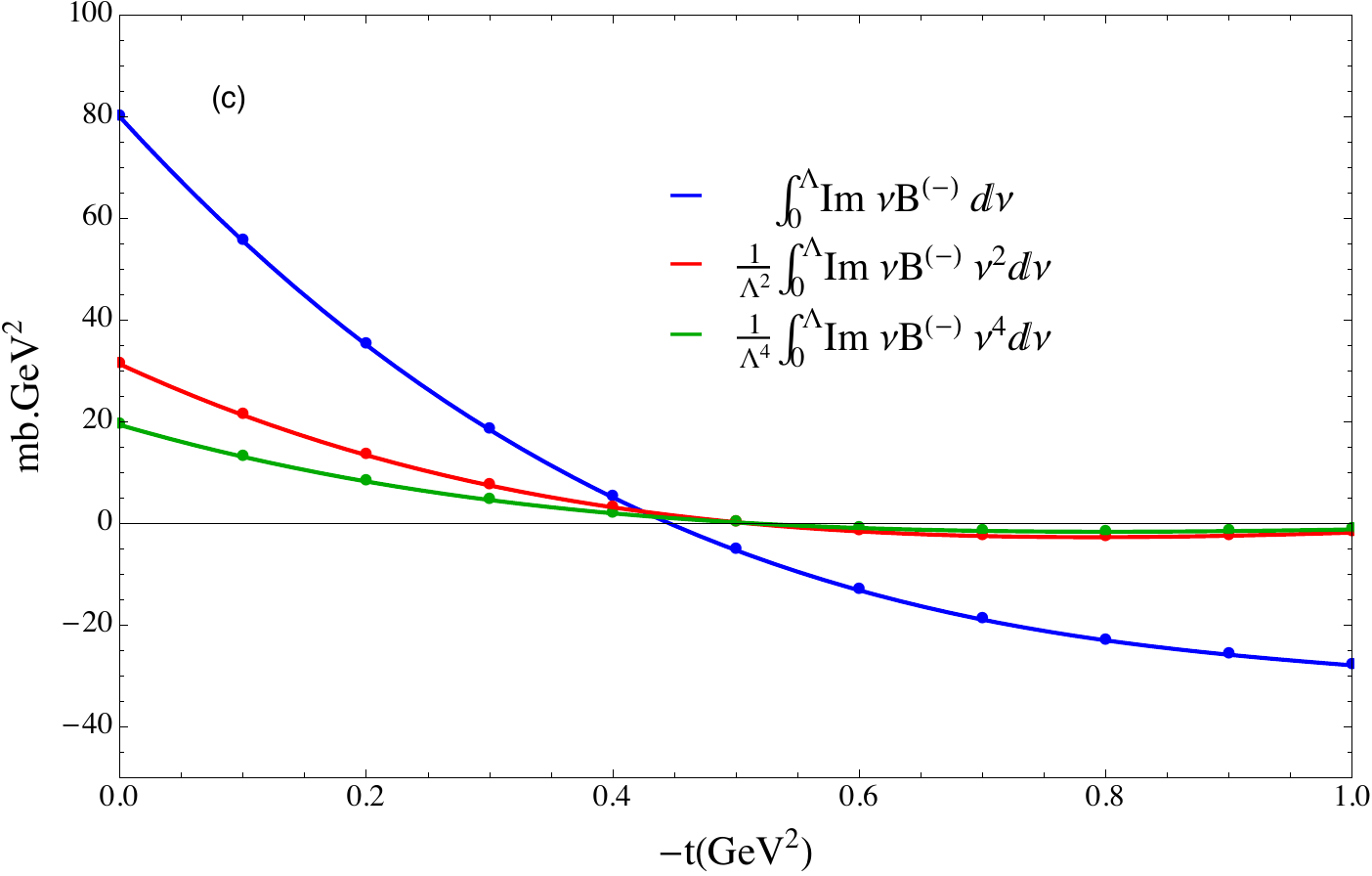}
 		\put (10,10) {\large$(c)$}
	\end{overpic}
	\begin{overpic}[width=0.45\textwidth]{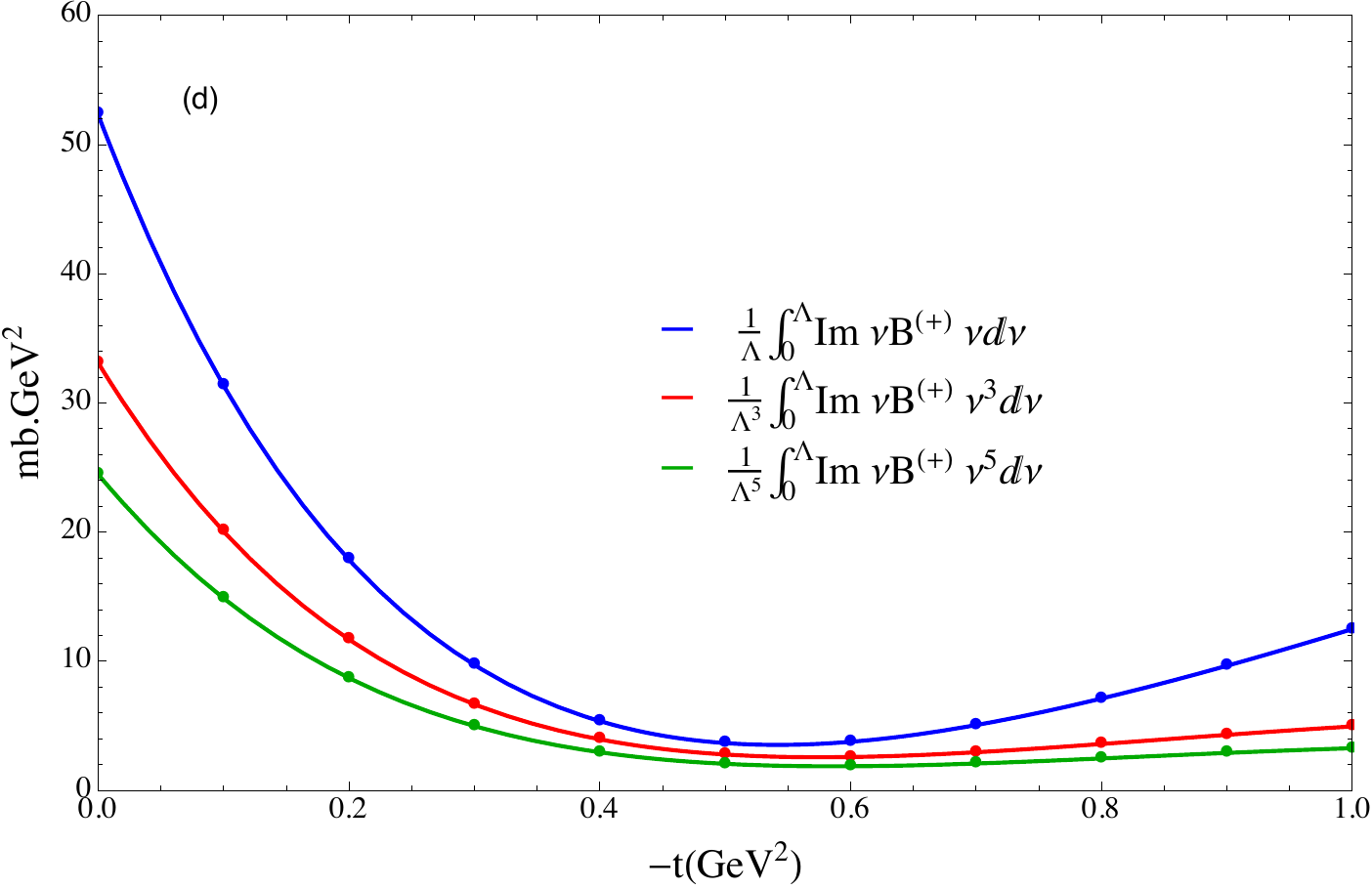}
 		\put (10,10) {\large$(d)$}
	\end{overpic}
  \caption{\label{fig:FESRSolt2} (color online) Left hand side of the sum rule, in Eq.\eqref{eq:SR},  computed for $k$ up to 5 and with $E^\text{max}_\text{lab}=2$ GeV . Fig (a) amplitude $A'^{(-)}$. Fig (b) amplitude $\nu B^{(-)}$. Fig (c) amplitude $A'^{(+)}$. Fig (d) amplitude $\nu B^{(+)}$. }
\end{center}
\end{figure*}

The results obtained for the left hand side of Eq.~(\ref{eq:SR})  for the two lowest moments, $S_k(\Lambda,t)$, $k=0,1$, using the four low-energy parametrizations (SAID, KH80, BoGa and Julich) of  for $A'^{(\pm)}$, and  $\nu B^{(\pm)}$  are shown in Fig. \ref{fig:FESRSolt}. All solutions yield similar {\it l.h.s.} for the sum rules. In Fig.~\ref{fig:FESRSolt2} we keep only the SAID model and show the left hand side of the sum rule for  higher moments, with $k$ up to $k=5$. Inspecting Figs.~\ref{fig:FESRSolt} and \ref{fig:FESRSolt2} we observe the following. 

\begin{itemize}
 \item The even moments $(k=0,2,4)$ of the crossing-odd helicity-flip amplitude $\nu B^{(-)}$ have a zero at  $t\sim -0.5$ GeV$^2$. If the right hand side of the sum rule were approximated by a single $\rho$ pole, this would imply a zero in the $\rho$ trajectory residue $\beta_\rho(t)$  at  $t\sim -0.5$ GeV$^2$. The $\rho$ trajectory function is approximated by $\alpha_\rho(t) \sim 0.5 + t$, which at $t \sim -0.5 \mbox{ GeV}^2$ yields $\alpha_\rho=0$, {\it i.e.} corresponds to an exchange of a particle with spin-0. For a helicity-flip amplitude this value of $t$ is referred to as a nonsense point since a particle of spin-0 cannot flip helicity at the nucleon vertex. Therefore the helicity-flip amplitude is expected to vanish at this point and this can be achieved, for example if $\beta_\rho(t)\propto \alpha_\rho(t)$ for $t$ near a nonsense value. This relation is referred to as the sense mechanism~\cite{collins} for inserting zero into the amplitude at a nonsense point. 

\item The 0-th moment of the crossing-odd helicity non-flip amplitudes $A'^{(-)}$ has a zero between  $t=0$ and $t= -0.1 \mbox{ GeV}^2$. The 2-nd and 4-th moments of the crossing-odd helicity non-flip amplitude $A'^{(-)}$ exhibit also a zero but it appears closer to the point $t=-0.1$ GeV$^2$. This is consistent with high-energy phenomenology where one observes a crossover between $\pi^+p$ and $\pi^-p$ differential cross sections as at $t\sim -0.1\mbox{GeV}^2$.  The crossover is attributed to the $\rho$ exchange since an isovector, $t$-channel exchange contributes with opposite signs to $\pi^+p$ and $\pi^- p$ amplitudes. The difference between differential cross sections for $\pi^- p$ and $\pi^+ p$ will therefore change sign if  the $\rho$ changes sign at $t\sim -0.1 \mbox{ GeV}^2$. 

\item The 1-st moment $(k=1)$ of the crossing-even helicity flip and non-flip amplitudes, $A'^{(+)}$ and  $\nu B^{(+)}$ have a  minimum at $t\sim -0.6$ GeV$^2$. These amplitudes involve exchange of the Pomeron and the $f_2$ pole. Both exchanges contribute significantly to the right hand side of the sum rule. The interpretation of the minimum is therefore not obvious. As will be shown in the following section we find that this minimum appears approximatively at the location of the signature-even zero of the $f_2$ trajectory at $\alpha_f=0$. There are theoretical reasons supporting the vanishing of the $f_2$ contribution at this point. However we haven't find a satisfactory explanation for the minimum of the right hand side of the sum rule. 

\item The odd moments ($k=1,3,5$) of the crossing-even $t-$channel helicity-flip amplitudes, $\nu B^{(+)}$  are quite similar to the corresponding moments of the $t-$channel non-flip amplitude $A'^{(+)}$. In other word the difference $A'-\nu B$ is  small. At large energies, the $s-$channel helicity-flip amplitude is proportional to the difference $A'-\nu B\approx A$ and only  the Pomeron and the $f_2$  contribute to the amplitude, $A^{(+)}=A^{\mathbb P}+A^f$, and analogously for $B$. The Pomeron is purely helicity non-flip in the $s-$channel, {\it i.e.} $\nu B^{\mathbb P} \gg A^{\mathbb P } \approx 0$. Thus the  residual contribution to $A'^{(+)}-\nu B^{(+)}\approx A^{(+)}\approx A^f$  originates from a small $s-$channel helicity flip contribution of the $f_2$ trajectory {\it i.e.} $\nu B^{f} \gg A^{f }  \neq 0$. 
Based on the above observations we conclude that the Pomeron and the $f_2$ contribute dominantly to the $s$-channel helicity non-flip amplitude, or, equivalently, that the isoscalar exchanges contribute equally to the $t-$channel helicity flip and non-flip. Hence, in the high energy region, we will use the same parametrization for $A'^{(+)}$ and $\nu B^{(+)}$. 

\end{itemize}


\subsection{High energy parametrization } \label{sec:high}
\begin{figure*}[htb]\begin{center}
	\begin{overpic}[width=0.3\textwidth]{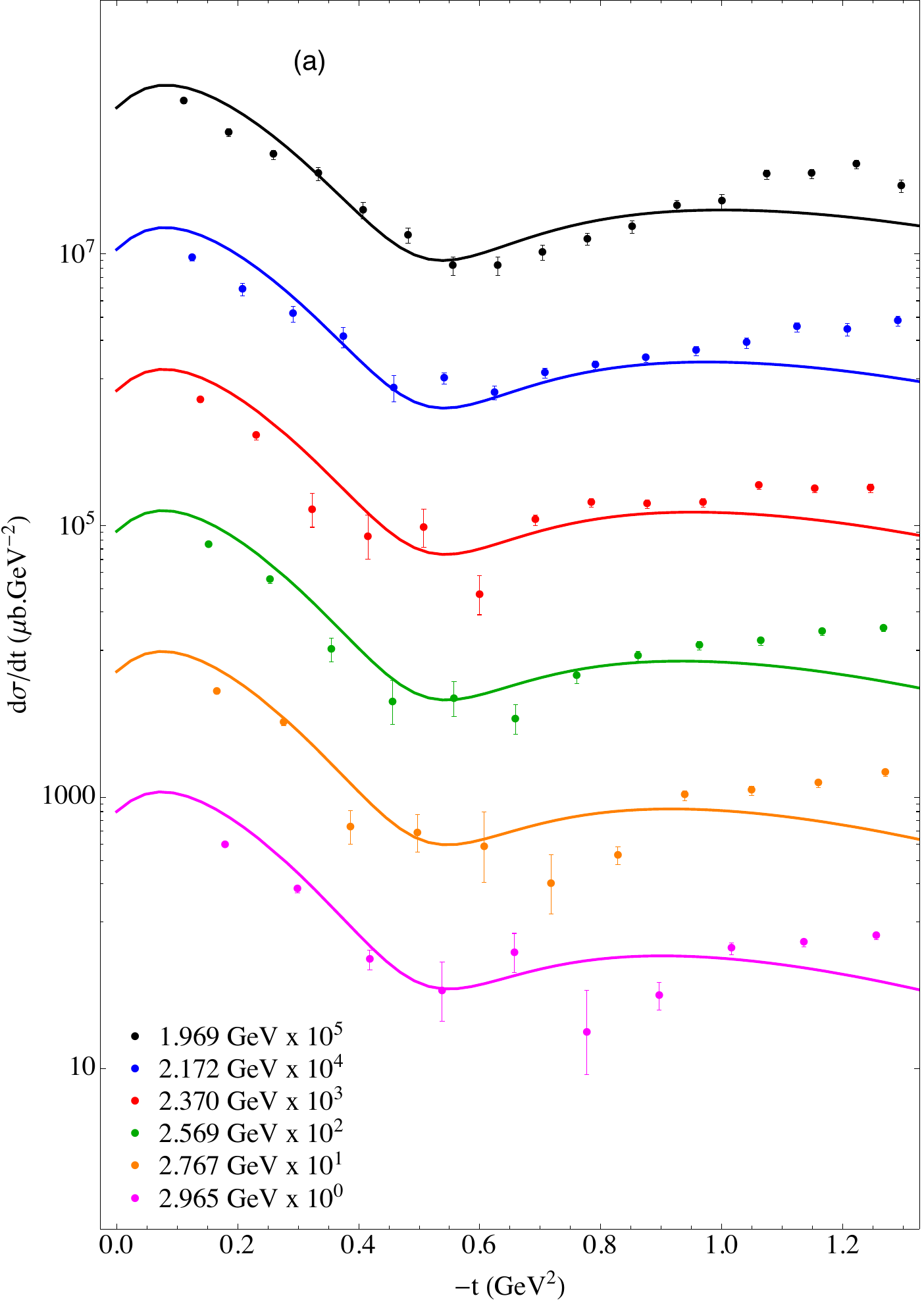}
 		\put (60,95) {\large$(a)$}
	\end{overpic}
	\begin{overpic}[width=0.3\textwidth]{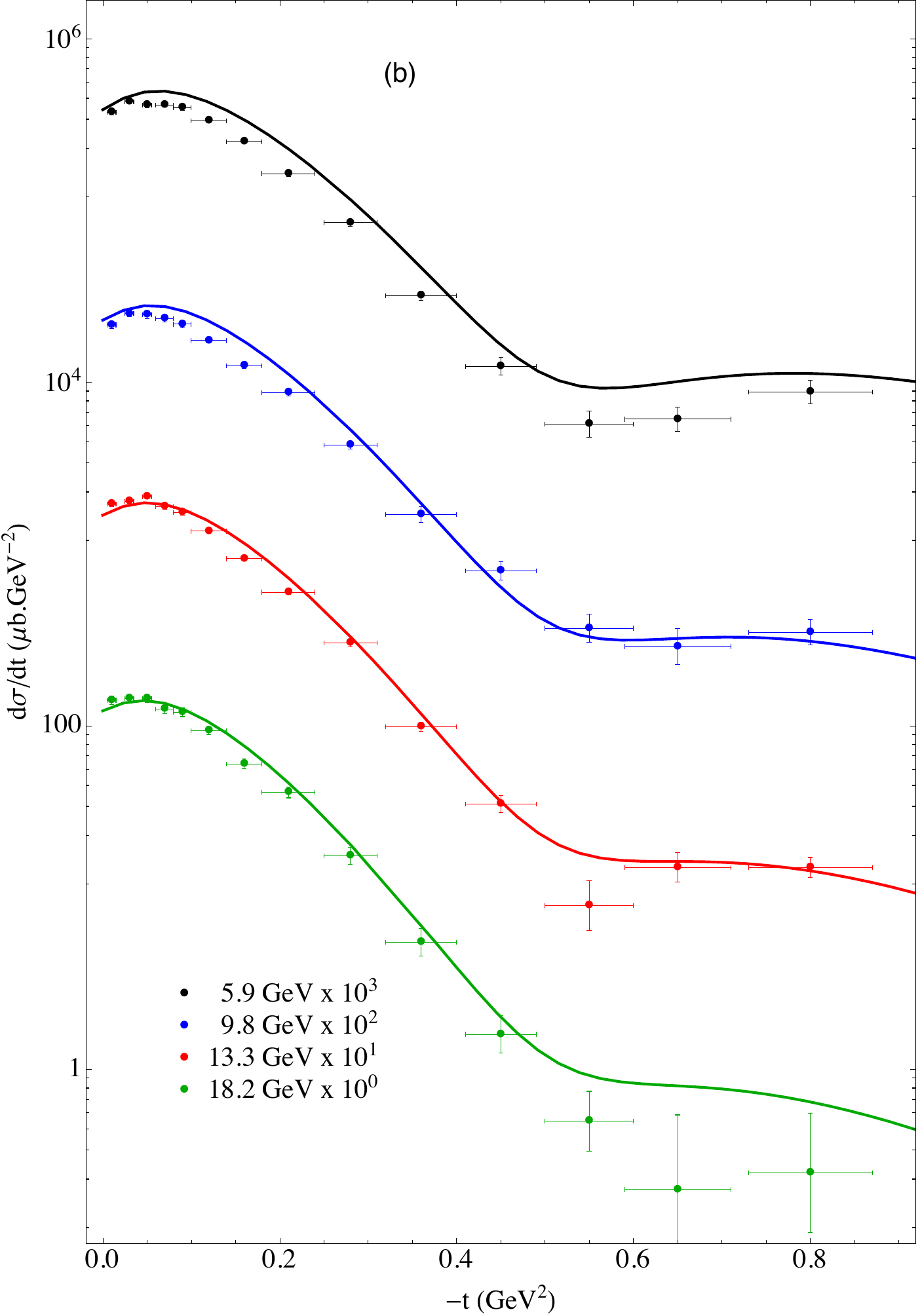}
 		\put (60,95) {\large$(b)$}
	\end{overpic}
	\begin{overpic}[width=0.3\textwidth]{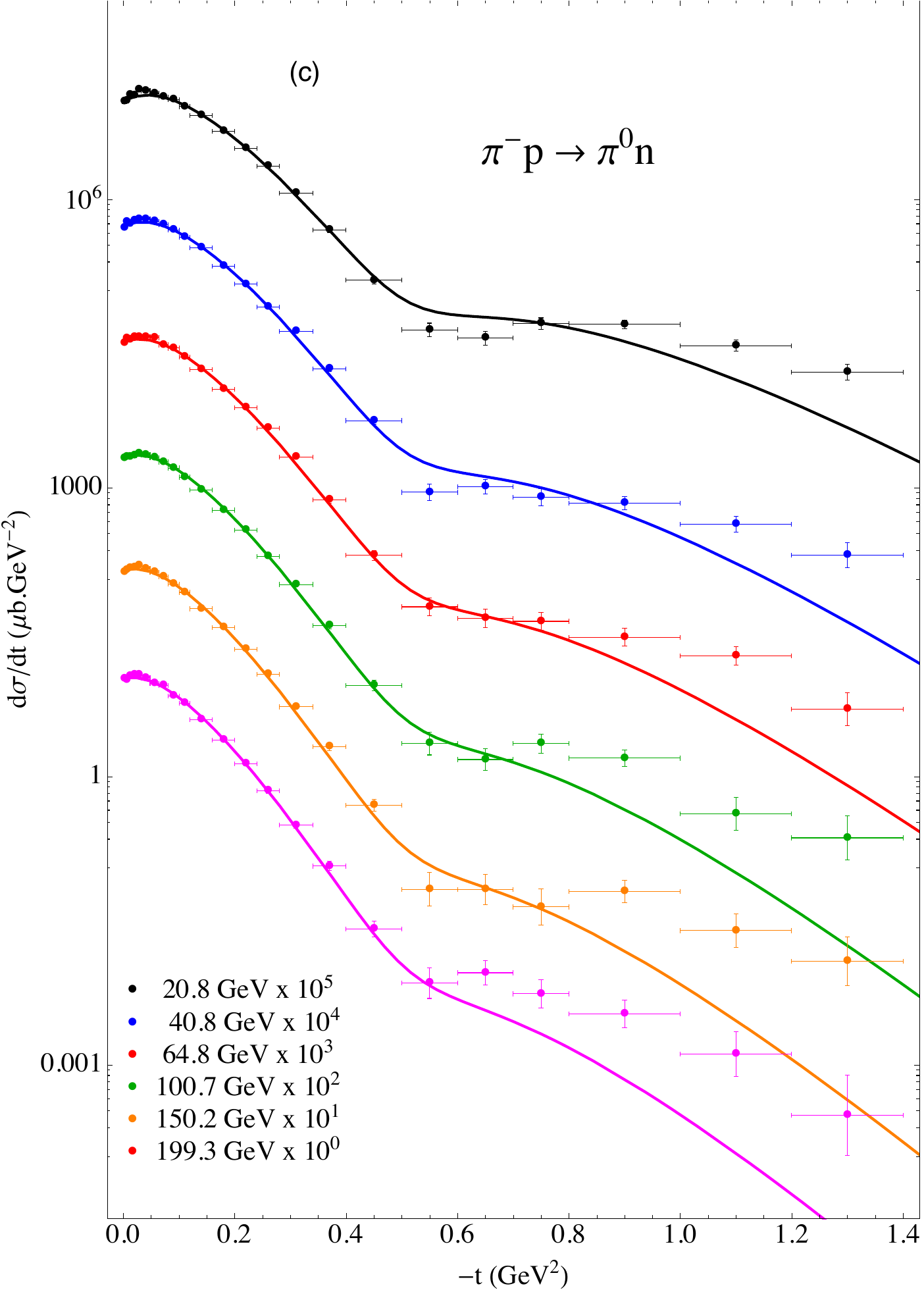}
 		\put (60,95) {\large$(c)$}
	\end{overpic}
\caption{\label{fig:PiNCEXSig}(color online) $\pi^- p\to \pi^0 n$ differential cross sections from $p_{\text{lab}}$=1.969 GeV to  $p_{\text{lab}}$=199.3 GeV. Scaling factors are indicated on the figure. The theoretical model includes the $\rho$ pole (solid line). Fig (a): data from \cite{Miyake}. Fig (b): data from \cite{Stirling:1965zz}. Fig (c): data from \cite{Barnes:1976ek}.}
\end{center}
\end{figure*} 

In this section we discuss parametrization of the $t-$channel helicity amplitudes Eq.~\eqref{eq:TCHA}  for $\pi N$ scattering in the high-energy region. As discussed in the preceding section, the leading asymptotic behavior of the $\pi N$ amplitudes involves three $t-$channel Regge poles $\mathbb P,f_2$ and $\rho$,  The first two have positive signature and contribute to the $t$-channel isoscalar amplitudes. The $\rho$ has negative signature and contributes to the isovector amplitudes. Schematically, displaying only factors originating from the isospin,  amplitudes for the three $\pi N$ reactions of interest are given by 
\begin{subequations}
\begin{align}
\pi^\mp p \to \pi^\mp p & = \mathbb P + f_2 \pm \rho, \\
\pi^- p \to \pi^0 n & = -\sqrt{2} \rho. 
\end{align}
\end{subequations}
Regge amplitudes in Eq.~\eqref{eq:Regge1} with even signature ($\tau = +1$)  have poles at even integer values of $\alpha$  while for odd signature ( $\tau=-1$) the poles occur for odd integer values of $\alpha$.  The poles corresponding to the reggeized $\rho$-exchange are physical if the pole is located at $\alpha_\rho(t) \ge 1$ and corresponds to a positive value of $t$. The unphysical poles located at  $\alpha_\rho\le 0$ ought to be canceled by residue zeros. 
To  remove such nonsense poles one choses the residue in the form  $\beta_\rho \propto 1/\Gamma(\alpha_\rho)$~\cite{Irving:1977ea}. In this case, for even (odd) integer $\alpha\le 0$,  the signature-odd amplitude vanishes (is finite). This pattern of residue zeros is consistent with FESR for the crossing-odd amplitudes helicity-flip amplitude $\nu B^{(-)}$, which, at high-energies, as discussed in the previous section,  is expected to have a zero at $\alpha_\rho=0$. 
As discussed in the previous section, the non-flip, isovector amplitude $A'^{(-)}$, however, is expected to be finite at $\alpha_\rho=0$, (corresponding to the point $t\sim -0.5 \mbox{ GeV}^2$).  
This is achieved by choosing $\beta_\rho \propto 1/\Gamma(\alpha+1)$. Furthermore, vanishing of $A'^{(-)}$ near $t=0$ is observed in the $\pi^\pm p$ crossover and we account for this by multiplying the residue by an additional factor $(1+C_2) e^{C_1 t}-C_2$, with $C_2$ chosen to reproduce the crossover in $\pi^\pm n$.  
 
With these parametrizations the two $t$-channel isovector amplitudes are predicted to vanish at the next nonsense wrong signature point, {\it i.e.} at   $\alpha_\rho=-2$,  which is located at $t\sim-2.8$ GeV$^2$. Unfortunately this point is beyond the range of applicability of our study since, as discussed earlier, truncation of the partial wave series prevents us from extrapolating the amplitudes   to such large values of $|t|$. 
        
The isovector $t-$channel amplitudes are therefore approximated by the $\rho$ Regge-pole and are given by\begin{subequations} \label{eq:Trho}
\begin{align}
A'^{(-)} &= \pi C^\rho_0\frac{\left[(1+C^\rho_2)e^{C^\rho_1 t}-C^\rho_2\right]}{\Gamma(\alpha_\rho+1)} \frac{e^{-i\pi \alpha_\rho}-1}{2 \sin\pi\alpha_\rho}   \nu^{\alpha_\rho}, \\  \label{eq:TrhoB}
B^{(-)} &= -D^\rho_0 e^{D^\rho_1 t} \frac{\pi}{\Gamma(\alpha_\rho)} \frac{e^{-i\pi \alpha_\rho}-1}{2 \sin\pi\alpha_\rho} \nu^{\alpha_\rho-1}.
\end{align}
\end{subequations}
The energy dependence is chosen such that the differential cross section behaves as  $d\sigma/dt\sim s^{2\alpha_\rho-2}$ at large energies. The relative sign is such that 
 the imaginary part of $A'^{(-)}$ and $B^{(-)}$ have the same sign as $C^\rho_0$ and $D^\rho_0$,  respectively.

In the following we use a linear trajectory for the $\rho$ pole, $\alpha_\rho=\alpha_\rho^0 + \alpha'_\rho t$. We first determine the parameters of the $\rho$ trajectories using only the data on the charge exchange   reaction $\pi^- p \to \pi^0 n$. Since the parameter $C^\rho_2$ is sensitive to the cross-over between $\pi^-p$ and $\pi^+p$ elastic scattering, our first fit cannot be used to determine $C^\rho_2$. We then impose the relation $C^\rho_2=\left[e^{0.1 C^\rho_1}-1\right]^{-1}$ such that the cross over arises at $t=-0.1$ GeV$^2$. 

At this stage, our model for the $t$-channel $\rho$ exchange involves six  parameters: magnitudes of the two residues, $C^\rho_0$ and $D^\rho_0$, two slope parameters, $C^\rho_1$ and $D^\rho_1$, and the intercept $\alpha_\rho^0$ and the slope $\alpha'_\rho$ of the $\rho$ trajectory. We fix these parameters  by fitting the differential cross section for the charge exchange reaction  $\pi^- p \to \pi^0 n$ using existing data for pion momentum, $p_L \ge 20\mbox{ GeV}$ \cite{Barnes:1976ek}. We extrapolate the model down to $p_\text{lab}=2$ GeV and compare it to the data on Fig.~\ref{fig:PiNCEXSig}. The results of the fit are summarized in the second column in Table~\ref{tab:parrho}. 

\begin{table}[h]\caption{Regge pole parameters. \label{tab:parrho}}\begin{center}
\begin{tabular}{|l|rrr|}
\hline
$x$ & $\rho\ \ \ \ \ \ \ $  & $\mathbb P\  \ \ \ \ \ \ $ & $f$\ \ \ \ \ \ \  \\\hline
$\alpha_x^0$ & $0.490\pm0.003$ & $1.075\pm0.001$ & $0.490$\\
$\alpha'_x$& $0.943\pm0.009$ &$0.434\pm0.002$ & $0.943$\\
$\alpha''_x$ & $-$ & $0.162\pm0.007$ & $-$\\
$C^x_0$ & $5.01\pm0.09$ & $23.89\pm0.09$ & $71.35\pm0.29$\\
$C^x_1$ & $10.10\pm0.21$ & $2.21\pm0.02$ & $3.18\pm0.04$\\
$D^x_0$ & $128.87\pm2.86$  & $-$ & $-$\\
$D^x_1$ & $1.38\pm0.07$ & $-$ & $-$\\
\hline
\end{tabular},\end{center}\end{table}

\begin{figure}[htb]\begin{center}   
	\includegraphics[width=\linewidth]{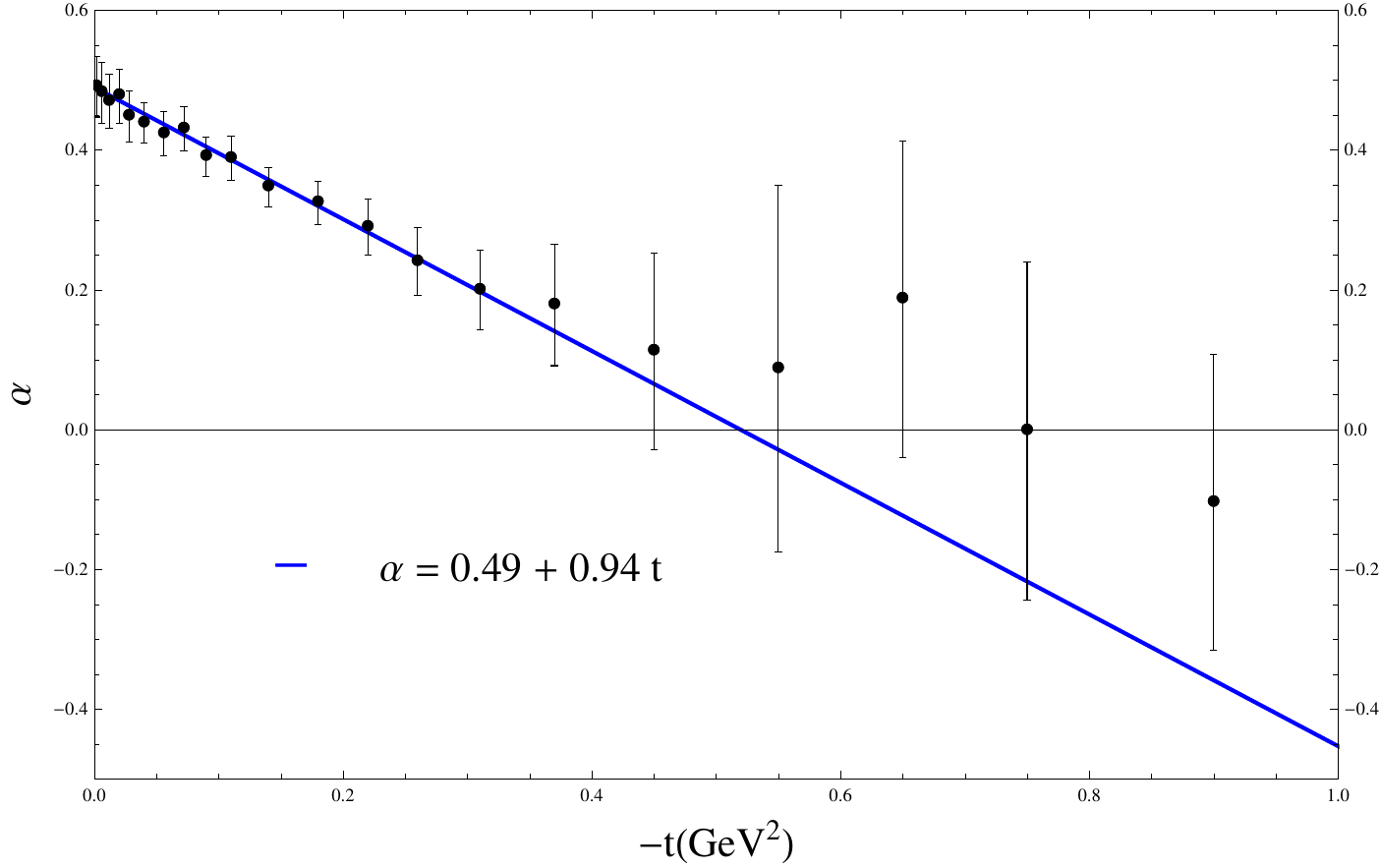} 
  \caption{\label{fig:alphacomp}(color online) $\rho$ trajectories from our model (blue solid line) and Barger and Phillips \cite{Barger:1970uz} (green dashed line) compared to effective trajectory extracted from data with Eq. \eqref{eq:trajectory}. We use data at $p_\text{lab}=20.8$ and 199.3 GeV from \cite{Barnes:1976ek}.}
\end{center}
\end{figure}

For $p_\text{lab}>20$~GeV corresponding to the energy range of the data in Ref.~\cite{Barnes:1976ek}, the $\rho$ pole dominates and we can neglect other contributions like Regge cuts and the $\rho'$ daughter trajectory. One can therefore assume a power law behavior for the energy dependence of the differential cross section and extract the $\rho$ trajectory from 
\be \label{eq:trajectory}
\alpha_\text{eff}(t) = \frac{1}{2} \log\left(\frac{p^2_a d\sigma(p_a,t)/dt}{ p^2_b d\sigma(p_b,t)/dt}\right) \log^{-1}\left(\frac{\nu_a}{\nu_b}\right).
\ee
We compare the effective trajectory extracted from the data~\cite{Barnes:1976ek} using $p_a=150.2$ GeV and $p_b=199.2$ GeV in Eq.~\eqref{eq:trajectory} and from our model in Fig.~\ref{fig:alphacomp}.  They clearly agree well as our trajectory is fitted to this data set.  The data support a linear trajectory at least up to the zero $\alpha_\rho(t)=0$. For the determination of the $\rho$ trajectory at higher $|t|$, we refer to the measurement of Refs~\cite{Barnes:1978kn,Barnes:1978fw} using semi-inclusive reaction. 

We now turn our attention to the isoscalar Regge poles. We assume that the isoscalar amplitudes are dominated by the Pomeron and the $f_2$ poles., {\it i.e.}
\begin{align} \label{eq:ABplus}
A'^{(+)} &= A'^{\mathbb P} + A'^{f}, & 
B^{(+)} &= B^{\mathbb P} + B^{f}.
\end{align}
The low-energy contribution to the FESR in Fig.~\ref{fig:FESRSolt} indicates that helicity-flip $\nu B^{(+)}$ and helicity non-flip  $A'^{(+)}$ isoscalar $t-$channel amplitudes are comparable. Phenomenologically the helicity non-flip amplitude $A'^{(+)}$, proportional to the total cross section, is more constrained by the data than the helicity flip amplitude $\nu B^{(+)}$. We choose to impose the equality between $t-$channel helicity flip and non-flip amplitudes in order to satisfy the FESR. The first physical particle on the $f_2$ trajectory is the $f_2(1275)$ spin-2 meson. To remove the ghost pole at $\alpha_f=0$ we use the parametrization
\begin{subequations} \label{eq:ABfP}
\begin{align} 
A'^{\mathbb P} &= -C_0^{\mathbb P}e^{C_1^{\mathbb P} t}\frac{\pi}{\Gamma(\alpha_{\mathbb P})} \frac{e^{-i\pi \alpha_{\mathbb P}}+1}{2 \sin\pi\alpha_{\mathbb P}}  \nu^{\alpha_{\mathbb P}}, & 
\nu B^{\mathbb P} &= A'^{\mathbb P},\\
A'^f &= -C_0^fe^{C_1^f t}\frac{\pi}{\Gamma(\alpha_f)} \frac{e^{-i\pi \alpha_f}+1}{2 \sin\pi\alpha_f}\nu^{\alpha_f}, & 
\nu B^f &= A'^f.
\end{align}
\end{subequations}
We choose the $f_2$ trajectory to be degenerate with the $\rho$, $\alpha_f = \alpha_\rho$. The degeneracy between the  $\rho$ and $f_2$ trajectories and residues follows from absence of exotic, isospin-2 mesons, {\it  e.g.}  in $\pi^+\pi^+$ scattering~\cite{Mandula:1970wz}. Degeneracy between the $f_2$ and $\rho$ and absence of ghost poles ($\alpha_f = 0$) is then consistent with the observed zero in the $\rho$ residue at $\alpha_\rho=0$ {\it cf.} Eq.(\ref{eq:Trho}b). 
   
\begin{figure*}[htb]\begin{center}   
	\begin{overpic}[width=\linewidth]{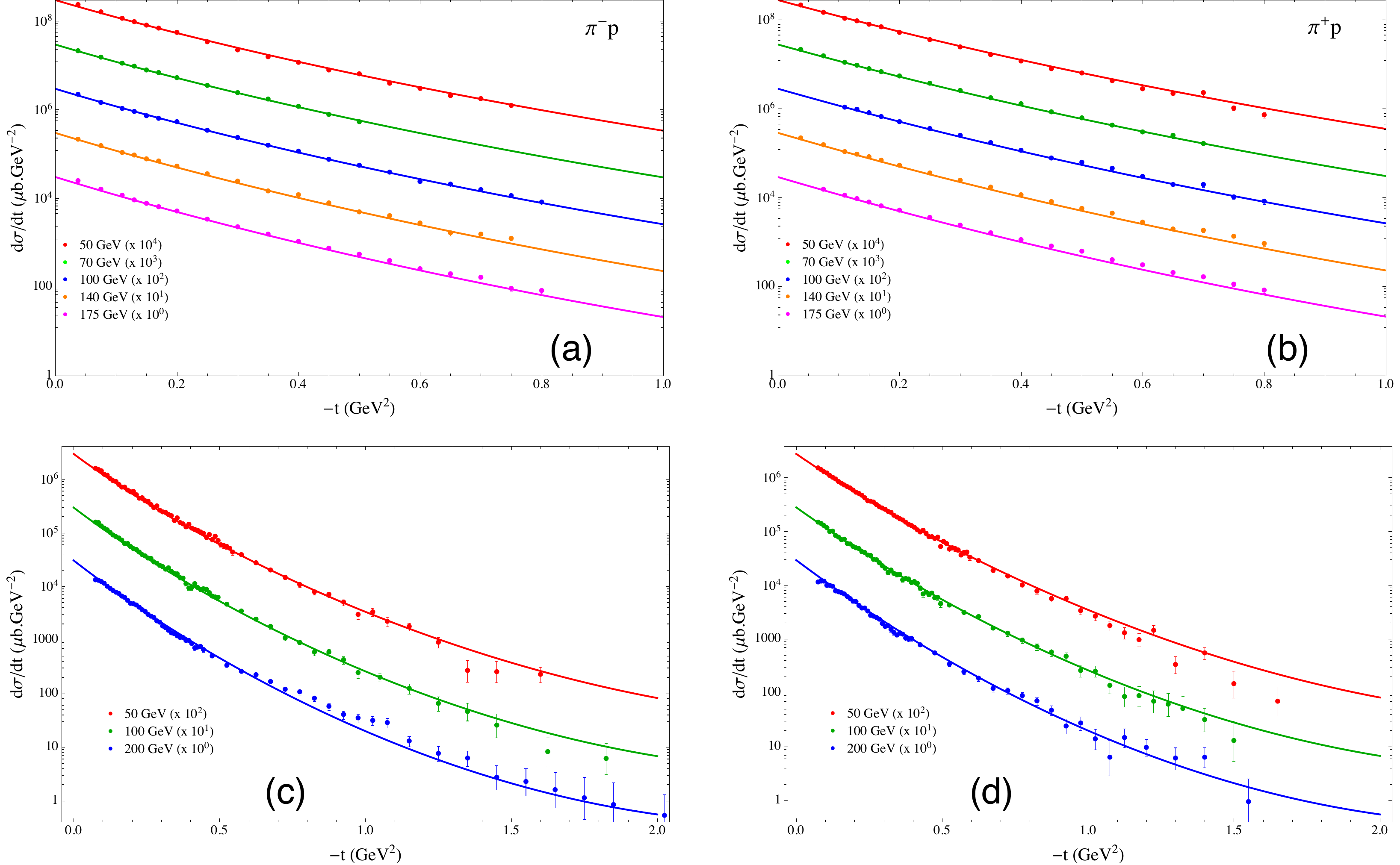}
	\put (5,36) {\large$(a)$}
	\put (5,5) {\large$(c)$}
	\put (60,36) {\large$(b)$}
	\put (60,5) {\large$(d)$}
	\end{overpic}
\caption{\label{fig:PiNSig1}(color online) $\pi^- p\to \pi^- p$ differential cross section for $p_{\text{lab}}\ge 50$ GeV compared to data from \cite{Ayres:1976zm} (a) and \cite{Akerlof:1976gk} (c). $\pi^+ p\to \pi^+ p$ differential cross section for $p_{\text{lab}}\ge 50$ GeV compared to data from \cite{Ayres:1976zm} (b) and \cite{Akerlof:1976gk} (d) The theoretical model (solid lines) includes the $\rho$, Pomeron and $f$ poles. The parameters are given in Table~\ref{tab:parrho}.  }
\end{center}
\end{figure*}

The Pomeron trajectory has a special status. There are no known mesons lying on it, with the exception that it may be related to the tensor glueball~\cite{Mathieu:2008me}.  
The trajectory is known to be approximately constant, $\alpha^{\mathbb P} \sim  1$.  In the following we parametrize it using a second order polynomial, 
\be
\alpha_{\mathbb P} = \alpha^0_{\mathbb P} + \alpha'_{\mathbb P} t + \alpha''_{\mathbb P} t^2,
\ee
to model the deviation from a straight line observed in the differential cross section {\it cf.} Fig.~\ref{fig:PiNSig1}. Over the range of $t$ considered here, the Pomeron trajectory is almost constant, and whether or not the factor $\Gamma(\alpha^{\mathbb P})$ is included is a matter of taste.

\begin{figure}[htb]\begin{center}
	\includegraphics[width=\linewidth]{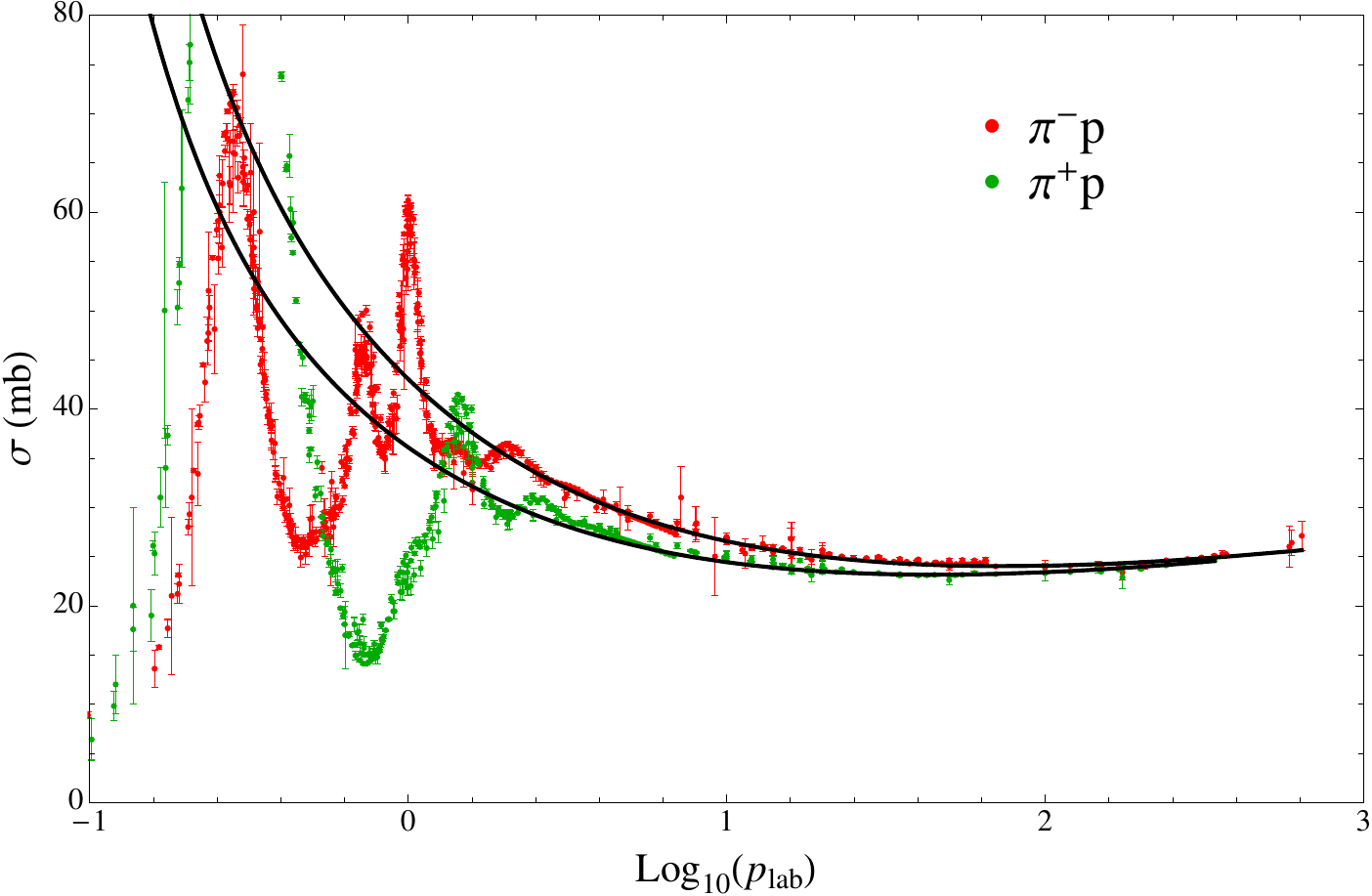}
  \caption{\label{fig:TotalSig}(color online) Total cross section. Data from \cite{Beringer:1900zz}.}
\end{center}
\end{figure}

In total we thus have seven parameters  describing the leading $t$-channel isoscalar Regge poles. Initially we attempted to fix these parameters, just like we did in the case of isovector exchanges, by  fitting the differential cross section. Since the Pomeron exchange, having the largest intercept, dominates and at the same time has a weak  $t$-dependent, we found that the error on the magnitude of the residue was large, of the order of 10$\%$. We therefore chose to perform a fit of the total cross sections (keeping only $p_\text{lab}\ge 5$ GeV data) to first  determine $C_0^{\mathbb P,f}$ and $\alpha^0_{\mathbb P}$ for the Pomeron.  The results are  shown in Fig.~\ref{fig:TotalSig}.  In the next step, using  the differential cross section for $p_\text{lab}>3$ GeV we determine the $f_2$ and Pomeron slope parameters $C_1^{\mathbb P,f}$, and the remaining Pomeron parameters that determine its $t$-dependence,   $\alpha'_{\mathbb P}$ and $\alpha''_{\mathbb P}$. The comparison with the data is shown in Fig.~\ref{fig:PiNSig1} for $p_\text{lab}\ge 50$ GeV. In the fit we use  the data from \cite{Ayres:1976zm,Akerlof:1976gk,Ambats:1974is}. The value of the parameters is given in columns three and four in Table~\ref{tab:parrho}.

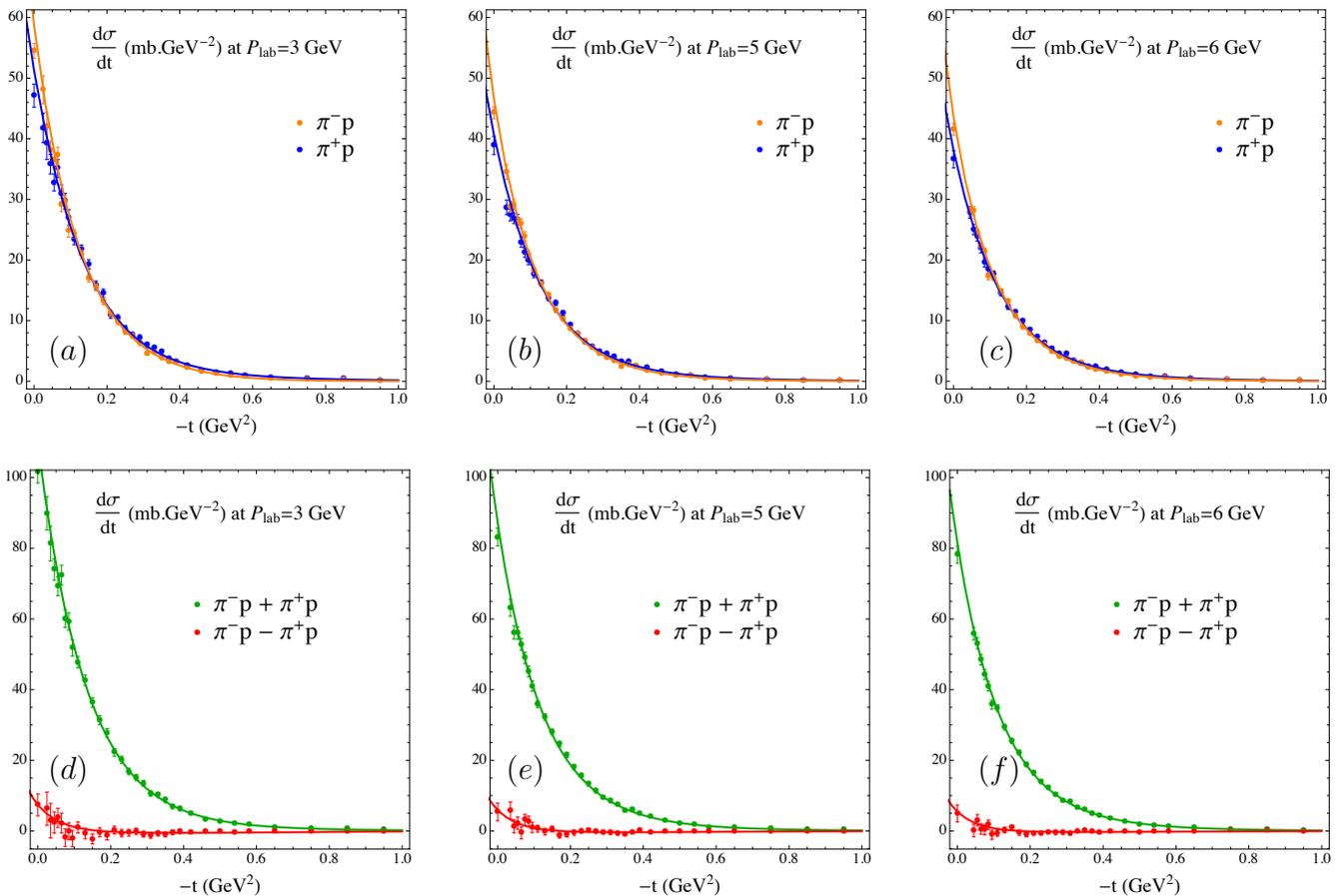
\begin{figure*}[htb]\begin{center}
	\begin{overpic}[width=\textwidth]{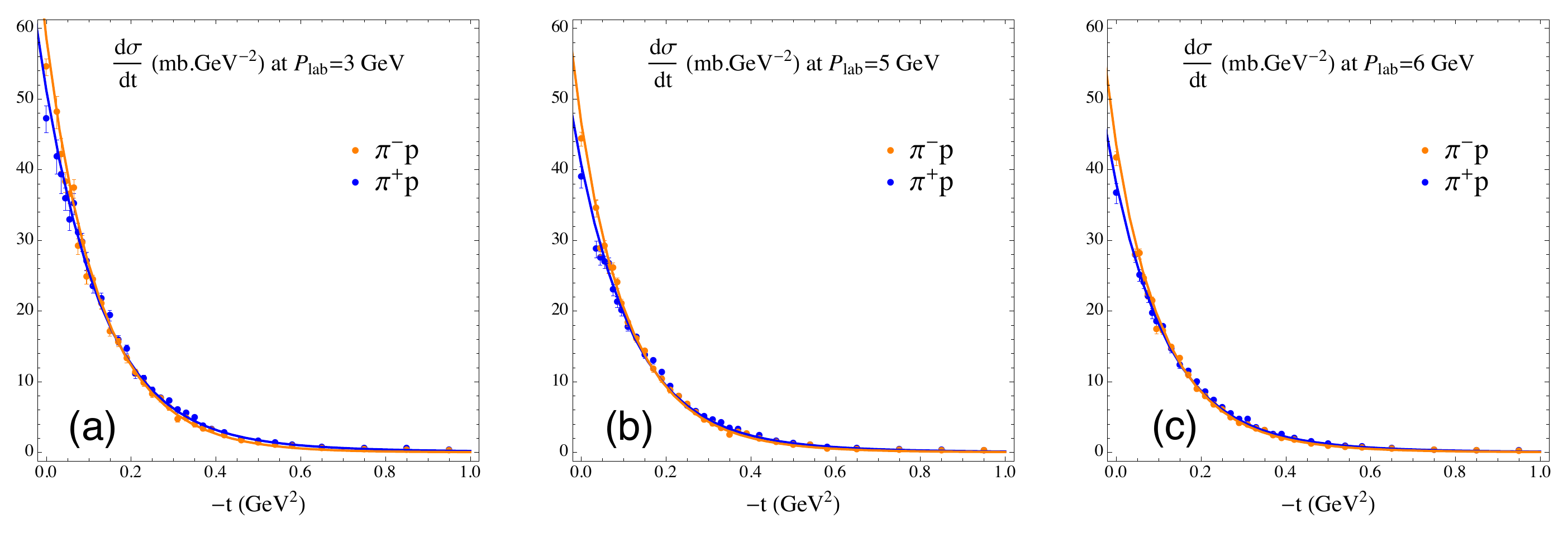}
 		\put (4,7) {\large$(a)$}
		\put (38,7) {\large$(b)$}
		\put (73,7) {\large$(c)$}
	\end{overpic}
	\begin{overpic}[width=\textwidth]{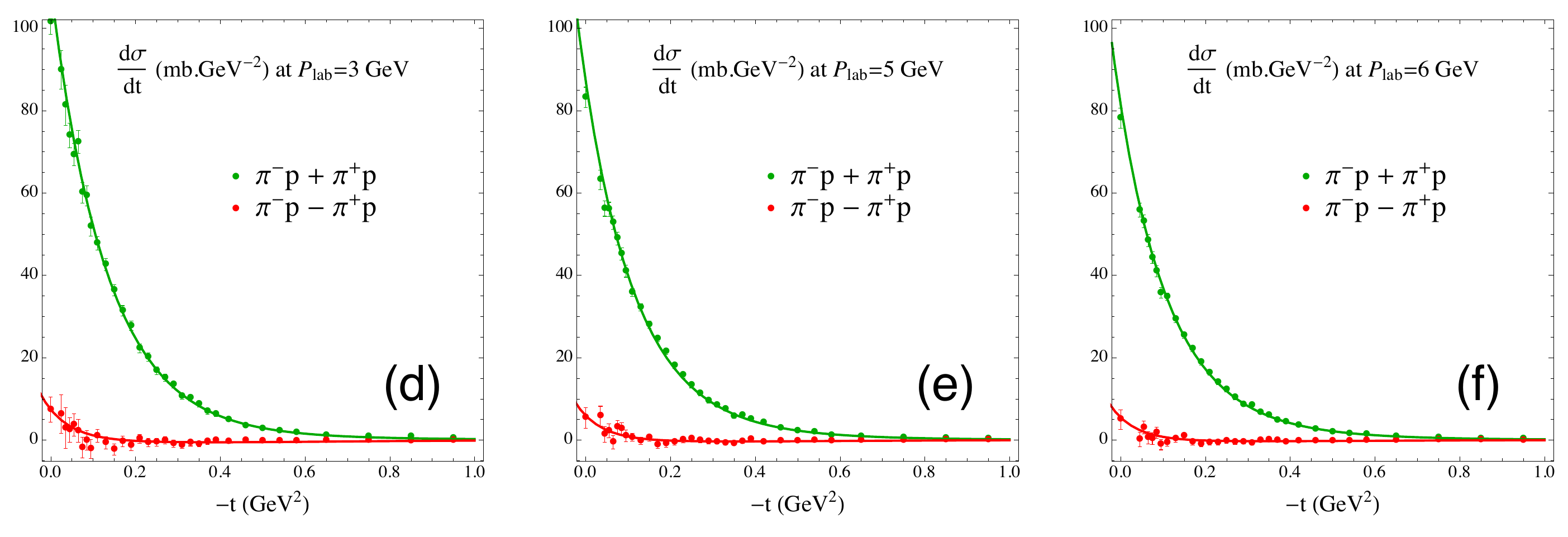}
 		\put (4,10) {\large$(d)$}
		\put (38,10) {\large$(e)$}
		\put (73,10) {\large$(f)$}
	\end{overpic}
 \caption{\label{fig:PiNSig2}(color online) $\pi^\pm p\to \pi^\pm p$ differential cross section for $p_{\text{lab}}$= 3 GeV (a), $p_{\text{lab}}$= 5 GeV (b)  and $p_{\text{lab}}$= 6 GeV (c).  The theoretical model (solid line) includes the $\rho$, Pomeron and $f$ poles. The data are from Ref.~ \cite{Ambats:1974is}. Fig (d), (e) and (f) by presenting the difference between $\pi^-p$ and $\pi^+ p$ emphases on the crossover at $p_{\text{lab}}$= 3,5,6 GeV respectively. }
\end{center}\end{figure*}

We compare our model with the differential cross section at $p_\text{lab}=3,5,6$ GeV from Ref.~\cite{Ambats:1974is} as shown on Fig.~\ref{fig:PiNSig2}. Our amplitudes reproduce the $\pi^\pm p$ differential cross section in whole range of $t$. 

\begin{figure*}[htb]\begin{center}    
	\begin{overpic}[width=0.49\textwidth]{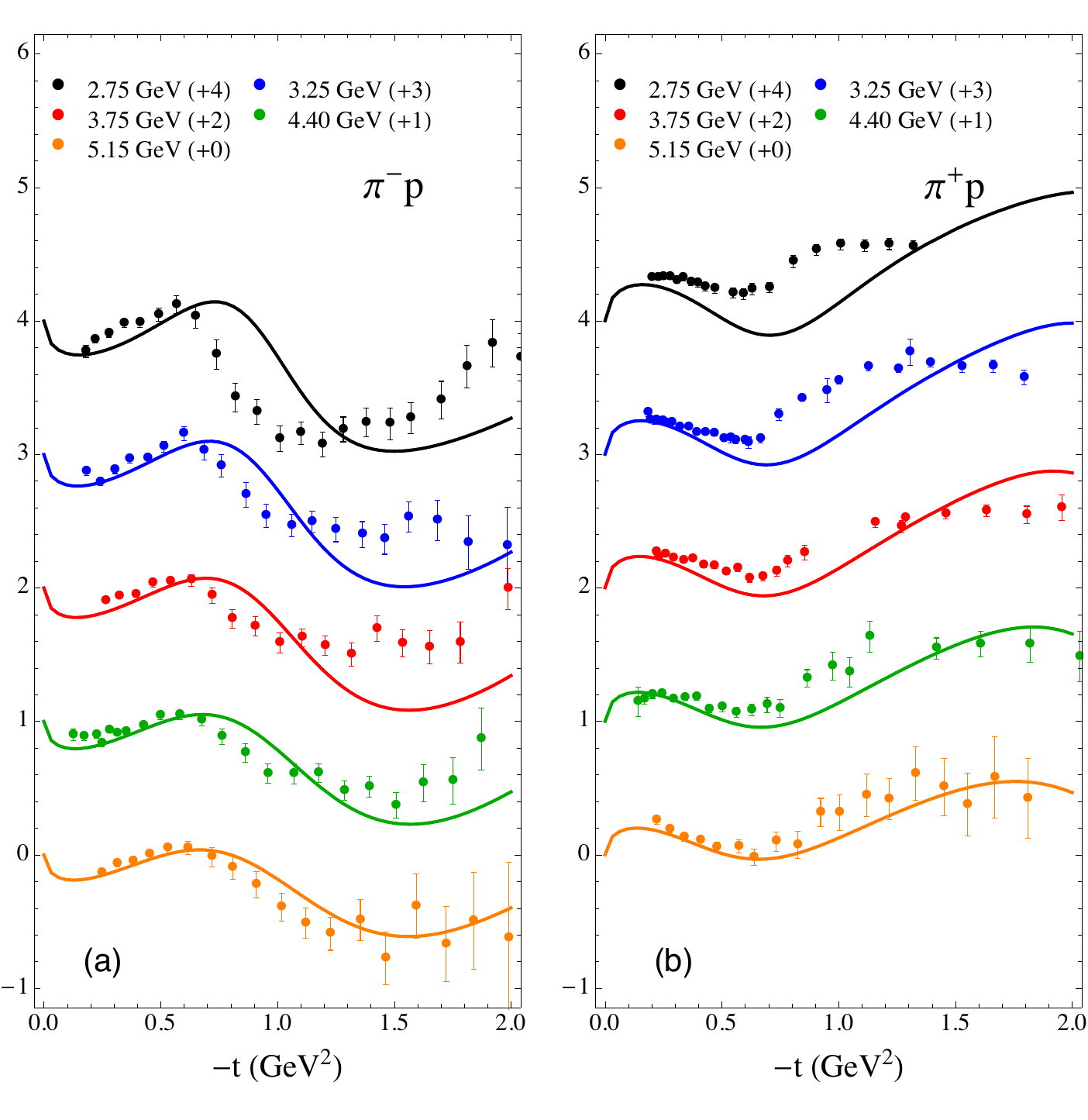}
 		\put (10,15) {\large$(a)$}
		\put (60,15) {\large$(b)$}
	\end{overpic}
	\begin{overpic}[width=0.49\textwidth]{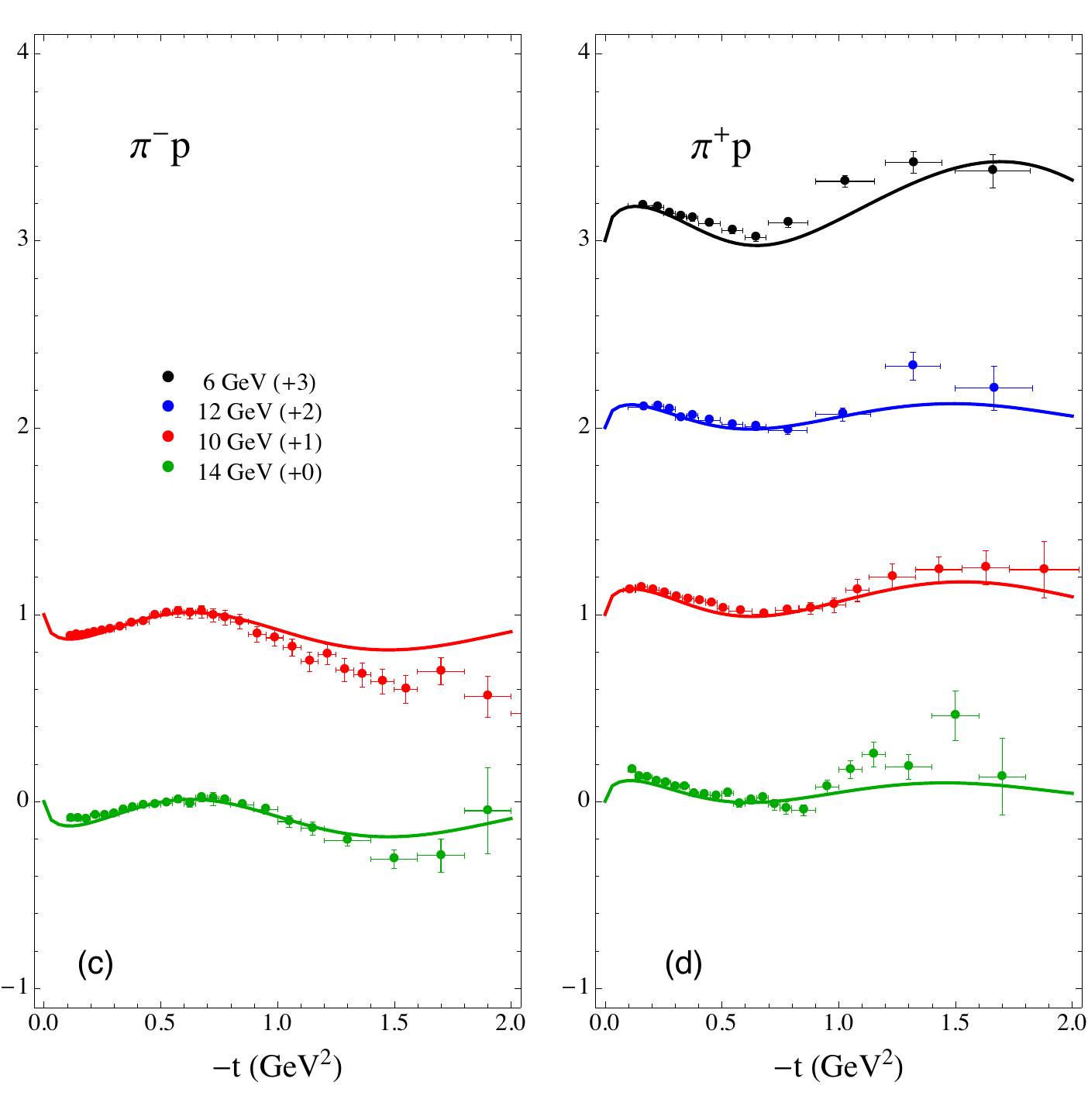}
 		\put (10,15) {\large$(c)$}
		\put (60,15) {\large$(d)$}
	\end{overpic}
  \caption{\label{fig:PiNPol}(color online) $\pi^\pm p\to \pi^\pm p$ polarization. The theoretical model (solid line) includes the $\rho$, Pomeron and $f$ poles. Fig (a) $\pi^- p\to \pi^- p$ polarization with data from \cite{Scheid:1973si}. Fig (b) $\pi^+ p\to \pi^+ p$ polarization with data from \cite{Scheid:1973si}. Fig (c) $\pi^- p\to \pi^- p$ polarization with data from \cite{Fujisaki:1979fm,Borghini:1971nu}. Fig (d) $\pi^+ p\to \pi^+ p$ polarization with data from \cite{Fujisaki:1979fm,Borghini:1971nu}.}
\end{center}
\end{figure*}

In the model the isovector contributions to the helicity non-flip amplitude is almost negligible.  If follows from Eq.~\eqref{eq:obsc}, that with the approximation $A'^{(-)}\approx 0$ polarizations in $\pi^+ p$ and $\pi^- p$ elastic scattering are predicted to be opposite to each other. This is  verified at  energies higher than $p_\text{lab}>5$ GeV, {\it cf.} as shown in Fig \ref{fig:PiNPol}.  
  
\subsection{Comparison between low- and high-energy contributions to the sum rules} 
\label{sec:comp}

\begin{figure*}[htb]\begin{center}   
	\begin{overpic}[width=0.49\textwidth]{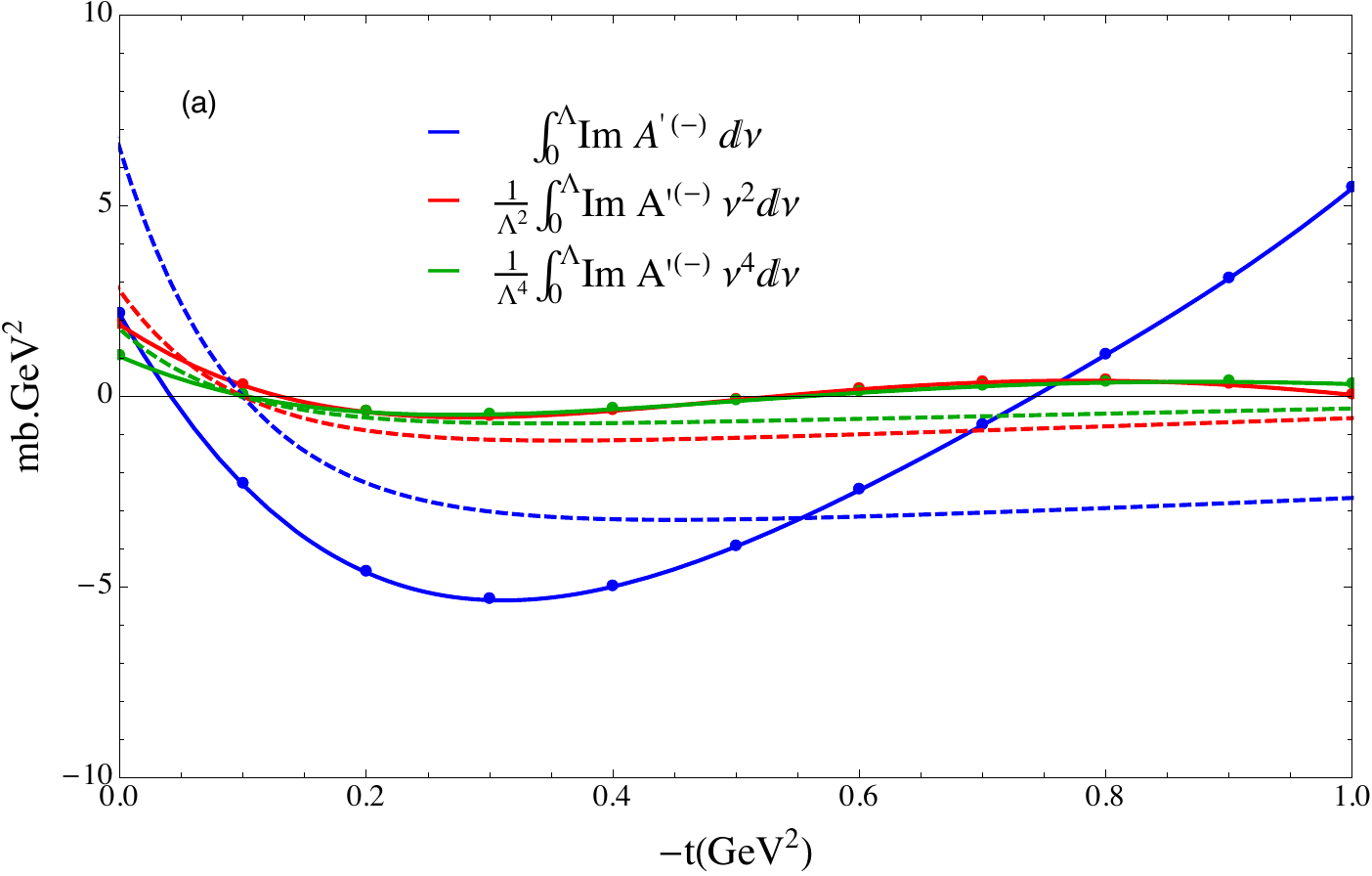}
 		\put (10,10) {\large$(a)$}
	\end{overpic}
	\begin{overpic}[width=0.49\textwidth]{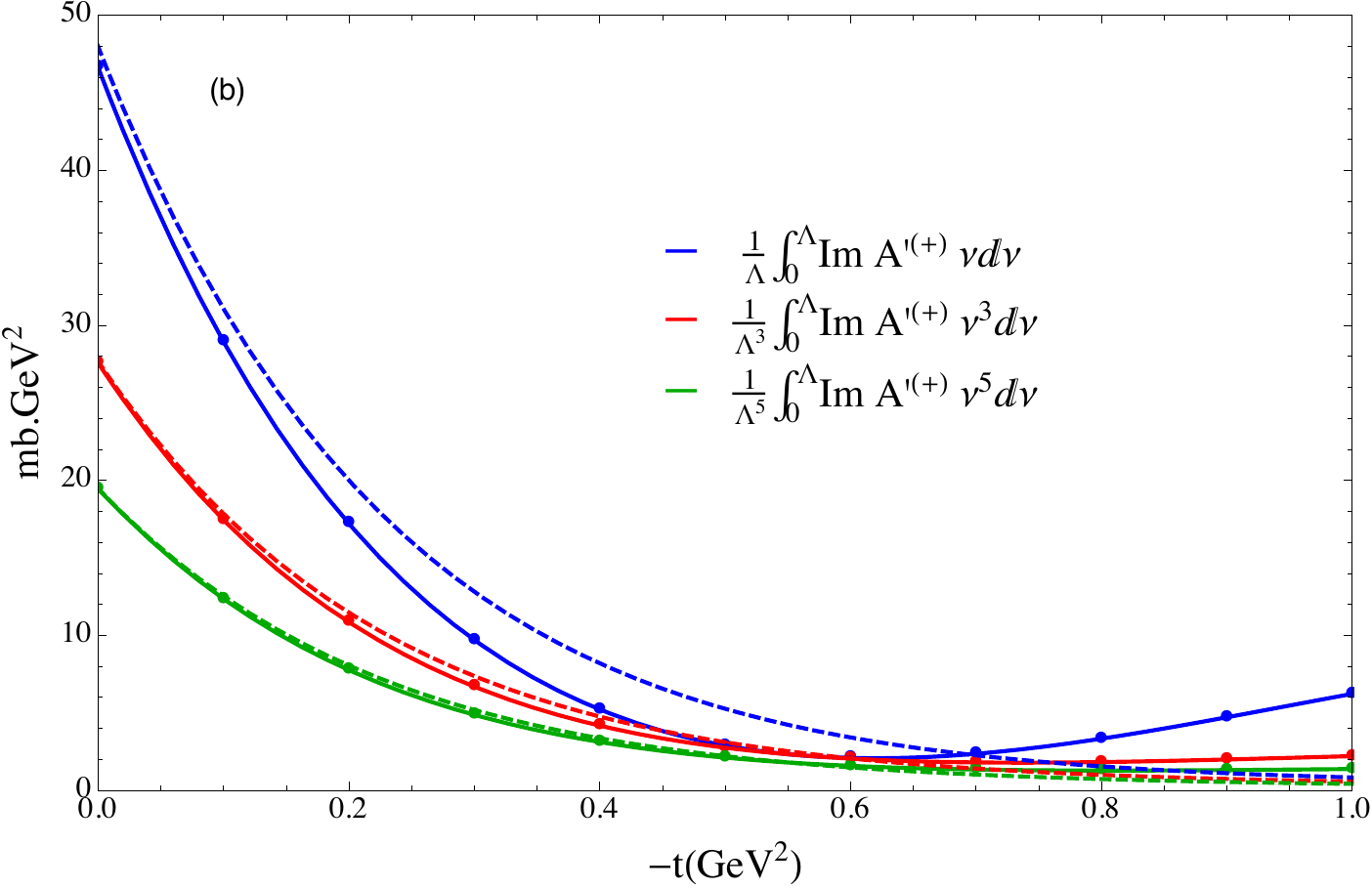}
 		\put (10,10) {\large$(b)$}
	\end{overpic}
	\begin{overpic}[width=0.49\textwidth]{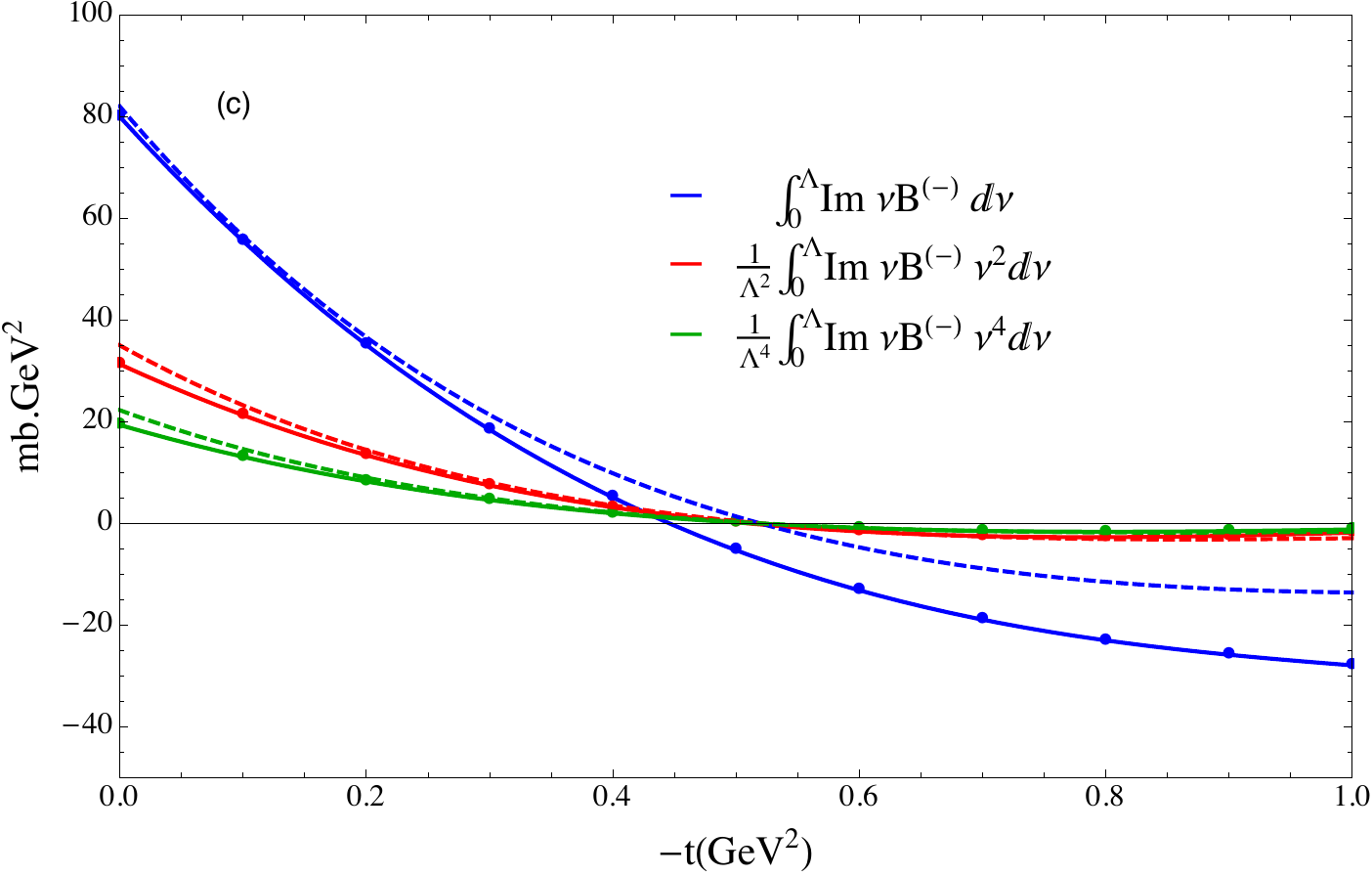}
 		\put (10,10) {\large$(c)$}
	\end{overpic}
	\begin{overpic}[width=0.49\textwidth]{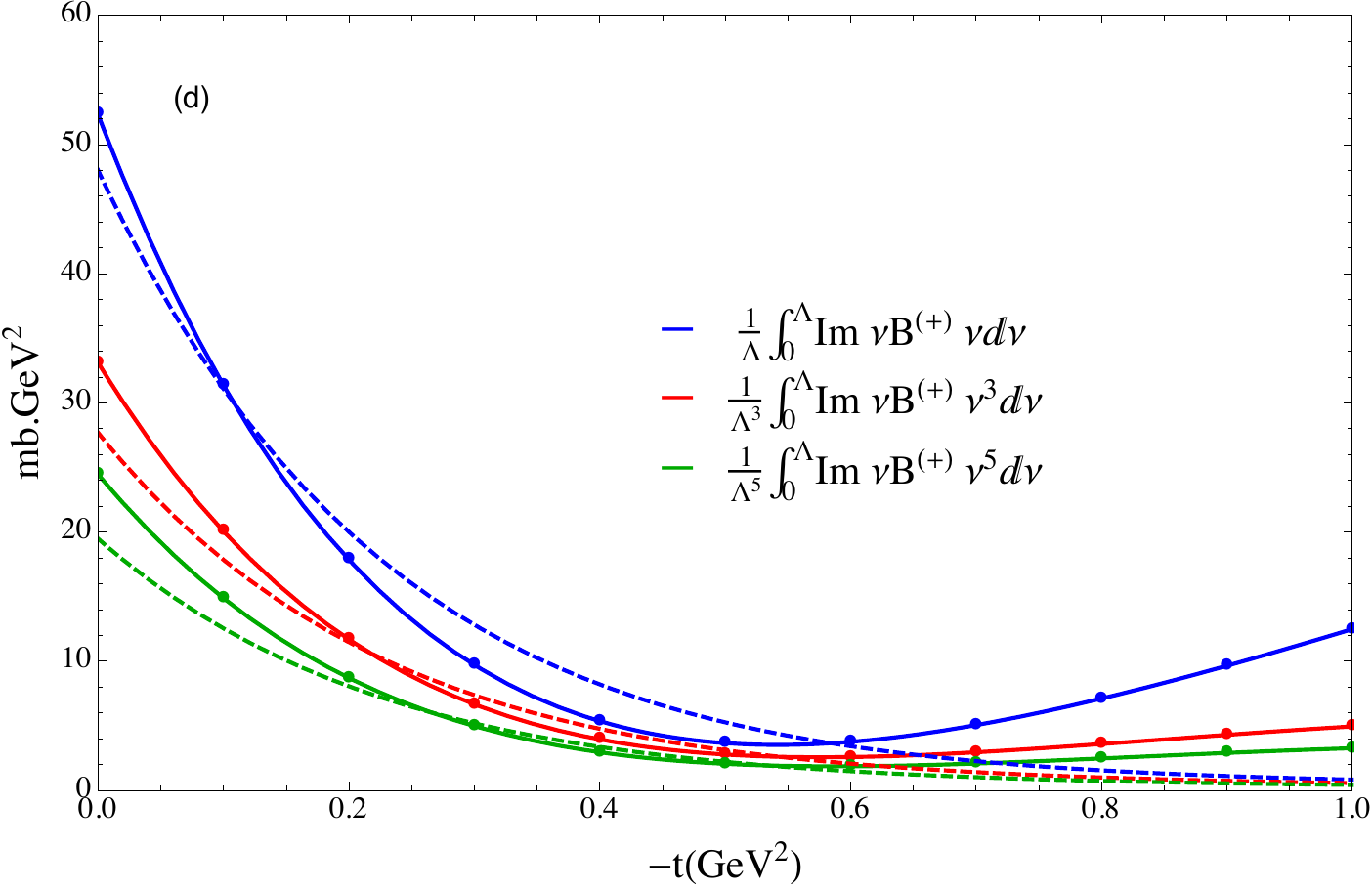}
 		\put (10,10) {\large$(d)$}
	\end{overpic}
	 \caption{\label{fig:FESRReg}(color online) Finite energy sum rules $S_k^\pm(\Lambda,t)$. Solid lines: left hand sides (low energy from SAID) ; dashed line: right hand sides (high energy). Fig (a) amplitude $A'^{(-)}$. Fig (b) amplitude $\nu B^{(-)}$. Fig (c) amplitude $A'^{(+)}$. Fig (d) amplitude $\nu B^{(+)}$.}
\end{center}
\end{figure*}

Having determined the parameters for the high energy model we can compute the right hand side of the sum rule in  Eq.~\eqref{eq:SR}. The comparison with the left hand side computed with the SAID solution, and discussed in Sec.~\ref{sec:low} is shown in Fig.~\ref{fig:FESRReg}. We compare the first three moments of the amplitudes $A'^{(\pm)}$ and $\nu B^{(\pm)}$. The same cutoff $\Lambda=E^\text{max}_\text{lab}+t/4M$ with $E^\text{max}_\text{lab}=2$ GeV is used in each sum rule. 

\begin{itemize} 
\item{}  The 0-th moment of the $t-$channel isovector, helicity non-flip amplitude, $A'^{(-)}$ changes sign at $t\sim -0.05$ GeV$^2$ but the 2-nd and 4th of this amplitude change sign at $t\sim -0.1$ GeV$^2$. As we explained, we included the change of sign at a fixed $t=-0.1$ GeV$^2$ in the parametrization~\eqref{eq:Trho}. The 2-nd and 4th moments of our model for the right hand side agree well with the left side of the sum rules. The 0-th moment of our model appears shifted at small $|t|$ compared to the 0-th moment of the SAID solution. This displacement might be caused by sub-leading Regge contributions ({\it e.g}  Regge cut or daughters trajectories). 

\item{} The moments of the $t-$channel isovector, helicity flip amplitude, $\nu B^{(-)}$ presents the same characteristic as the non-flip amplitude $A'^{(-)}$: the left hand side of the sum changes sign but for the lowest moment, the crossing point appears at a smaller value of $|t|$ than for all the other moments. In our model for the high energy region of this amplitude, we included only the dominant $\rho$ pole with a residue vanishing at the non-sense point $\alpha_\rho=0$. Thus, the crossing for the right hand side of the sum rule appears at the same $|t|$ for all moments. And the crossing point, given by the $\rho$ trajectory, $t=-0.52$ GeV$^2$ is in agreement with the 2-nd and 4-th moments of the SAID solution. As in the non-flip amplitude, a sub-leading Regge singularity whose influence would be non-negligible only in the 0-th moment, could be responsible for this deviation. 

\item{} The sum rules for the $k=1,3,5$ moments of $A'^{(+)}$ are well satisfied. The high energy parametrization at $t=0$ is largely constrained by the total cross section. 

\item{} As we explained before, {\it cf.} Eqs~\eqref{eq:ABplus} and~\eqref{eq:ABfP}, we imposed the condition $\nu B^{(+)}=A'^{(+)}$ at high energy. We thus have no freedom in the high energy parametrization of $\nu B^{(+)}$ and the sum rules for the $k=1,3,5$ moments are only qualitatively satisfied. The difference $A'-\nu B$ is, at high energy, approximatively the $s-$channel helicity flip amplitude. The isoscalar exchanges have small $s-$channel helicity flip amplitude at hight energies~\cite{Irving:1977ea}. We have neglected this contribution since the data are not very sensitive to it. 
\end{itemize}  

\begin{figure}[htb]\begin{center} 
	\includegraphics[width=\linewidth]{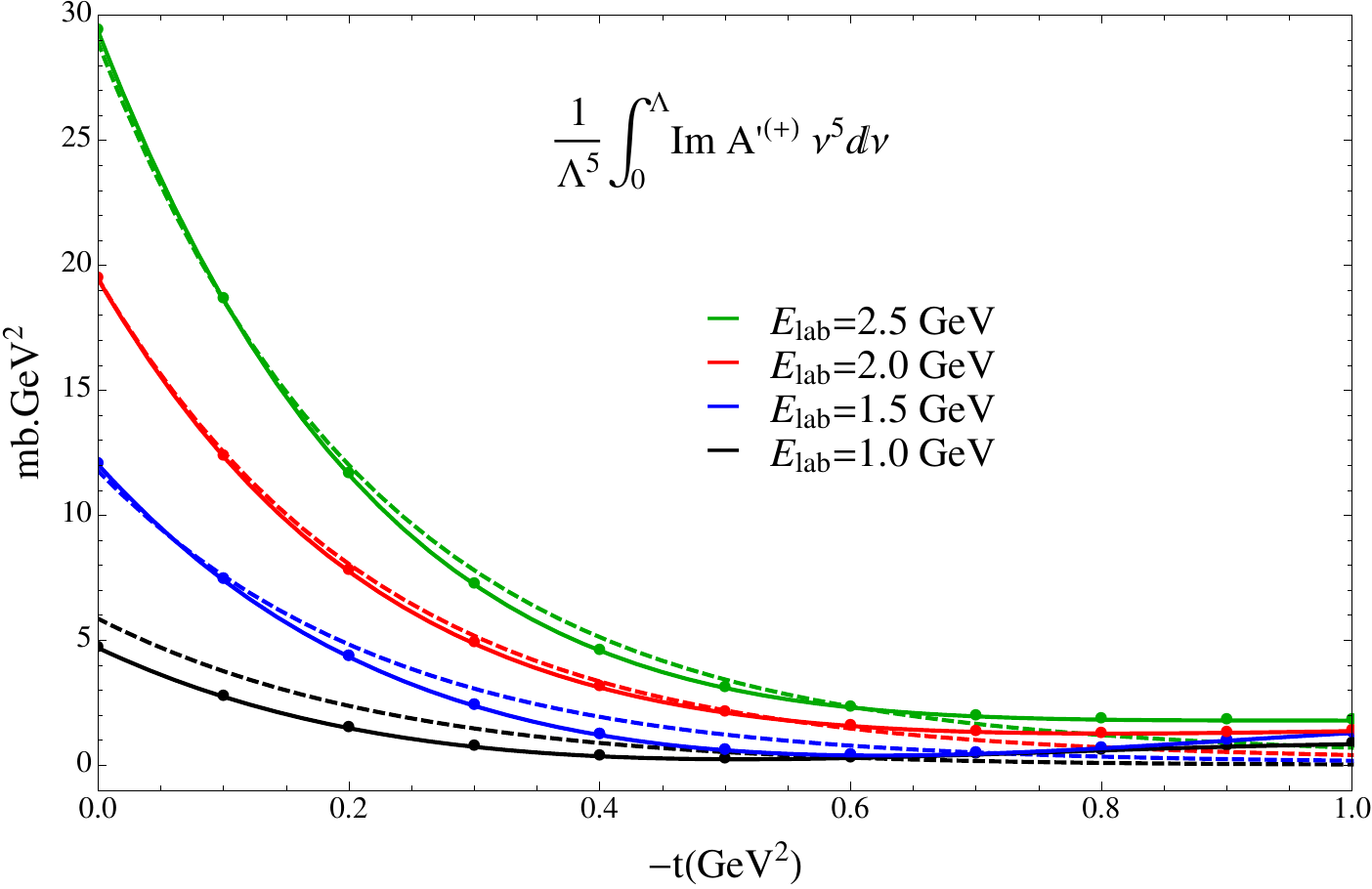} 
  \caption{\label{fig:FESRcutoff}(color online) Both side of the sum rule for the 5-th moment of $A'^{(+)}$. The solid (dashed) line is the right (left) hand side of the sum rule. The cutoff is $\Lambda=E_\text{lab}+t/4M$. }
\end{center}
\end{figure}

The results discussed above correspond to  fixed $\Lambda=E^\text{max}_\text{lab}+t/4M$ with $E^\text{max}_\text{lab}=2$ GeV. We have also investigated sensitivity to variations in $\Lambda$. The total cross sections shown in Fig. \ref{fig:TotalSig} shows resonance behavior  up to $E_\text{lab}\sim 1.6-2$ GeV. At higher energies the total cross section is smooth and well described by a sum of Regge poles. The range  $E_\text{lab}\sim 1.6-2$ GeV corresponds to the transition region. The cutoff in energy sum rules should be chosen in that region. We use $A'^{(+)}$, the amplitude that seems to best satisfy the FESR to study $\Lambda$ dependence. In Fig.~\ref{fig:FESRcutoff} we compare both sides of the sum rule for the $k=5$ moment when $E^\text{max}_\text{lab}$ takes the values of $1$ GeV, $1.5$ GeV, $2$ GeV and $2.5$ GeV. Near the forward direction, the FESR is satisfied only for $E_\text{lab}\ge 1.5$ GeV, which confirms the transition between resonances and Regge pole observed in the total cross section.

\section{The New Amplitudes} \label{sec:new}

One can contemplate the following strategy for an improved partial wave analysis that incorporates the high-energy data. A model is proposed, {\it i.e.} as in the SAID model, for the imaginary part of the amplitudes below $E_\text{lab}\sim 2$ GeV. In this energy range, the model can contain a limited number of  partial waves.  A different model,  based on Regge exchanges is constructed for $E_\text{lab}> 2$ GeV. The parameters of the high-energy model are constrained by two  independent conditions. One is the high energy data itself, the other is the FESR. The imaginary part of the amplitudes in the whole physical region $\nu\in[\nu_0,\infty[$ is obtained by  interpolating between the imaginary part of the partial wave series at low energies and the imaginary part of the Regge model at high energies. In the intermediate energy range, different techniques, {\it e.g.} linear interpolation or conformal mapping can be used to match the two models. Real parts of the amplitudes are then reconstructed using dispersion relations.

\begin{figure*}[htb]\begin{center} 
	\begin{overpic}[width=0.49\textwidth]{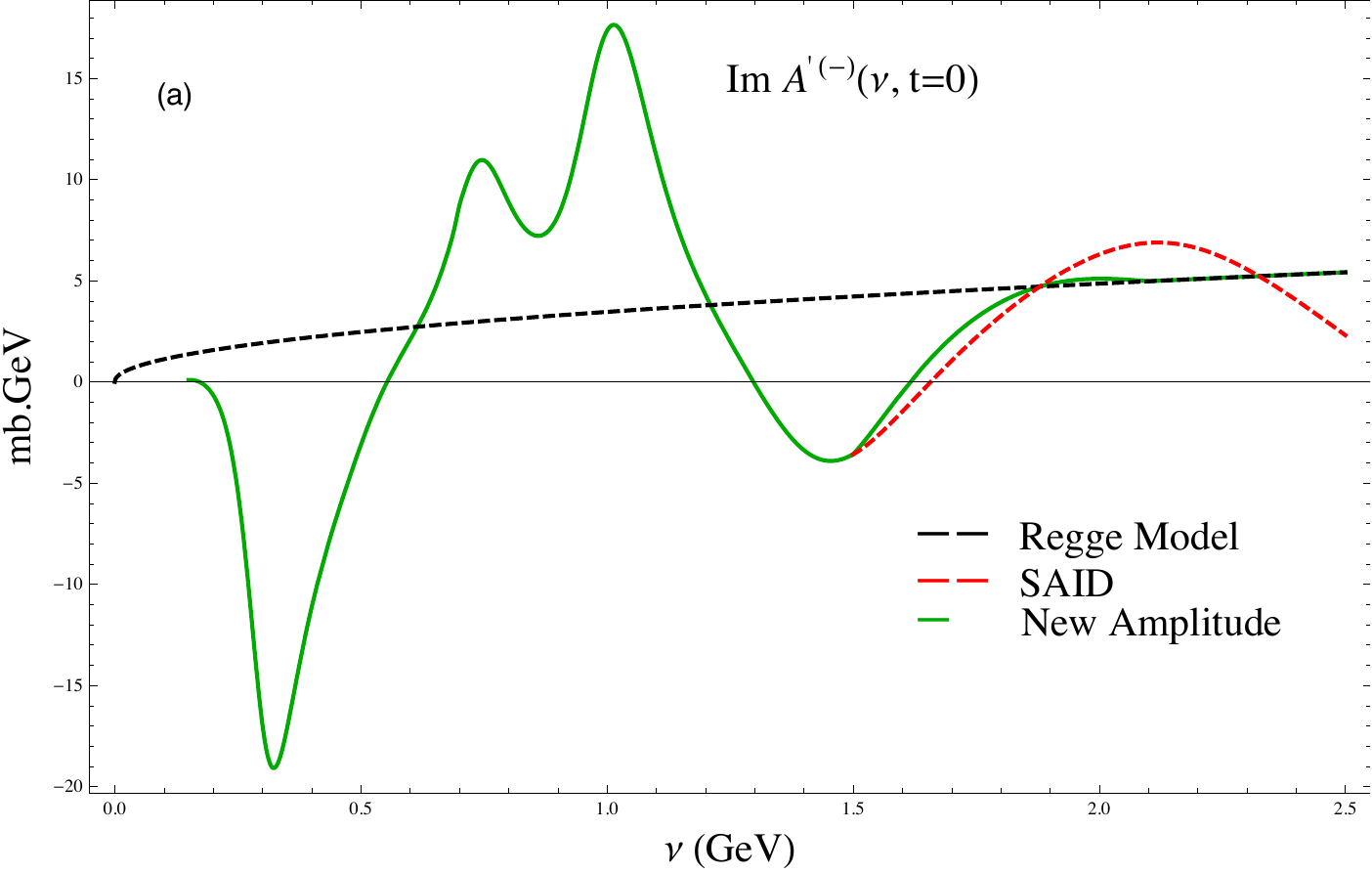}
 		\put (90,10) {\large$(a)$}
	\end{overpic}
	\begin{overpic}[width=0.49\textwidth]{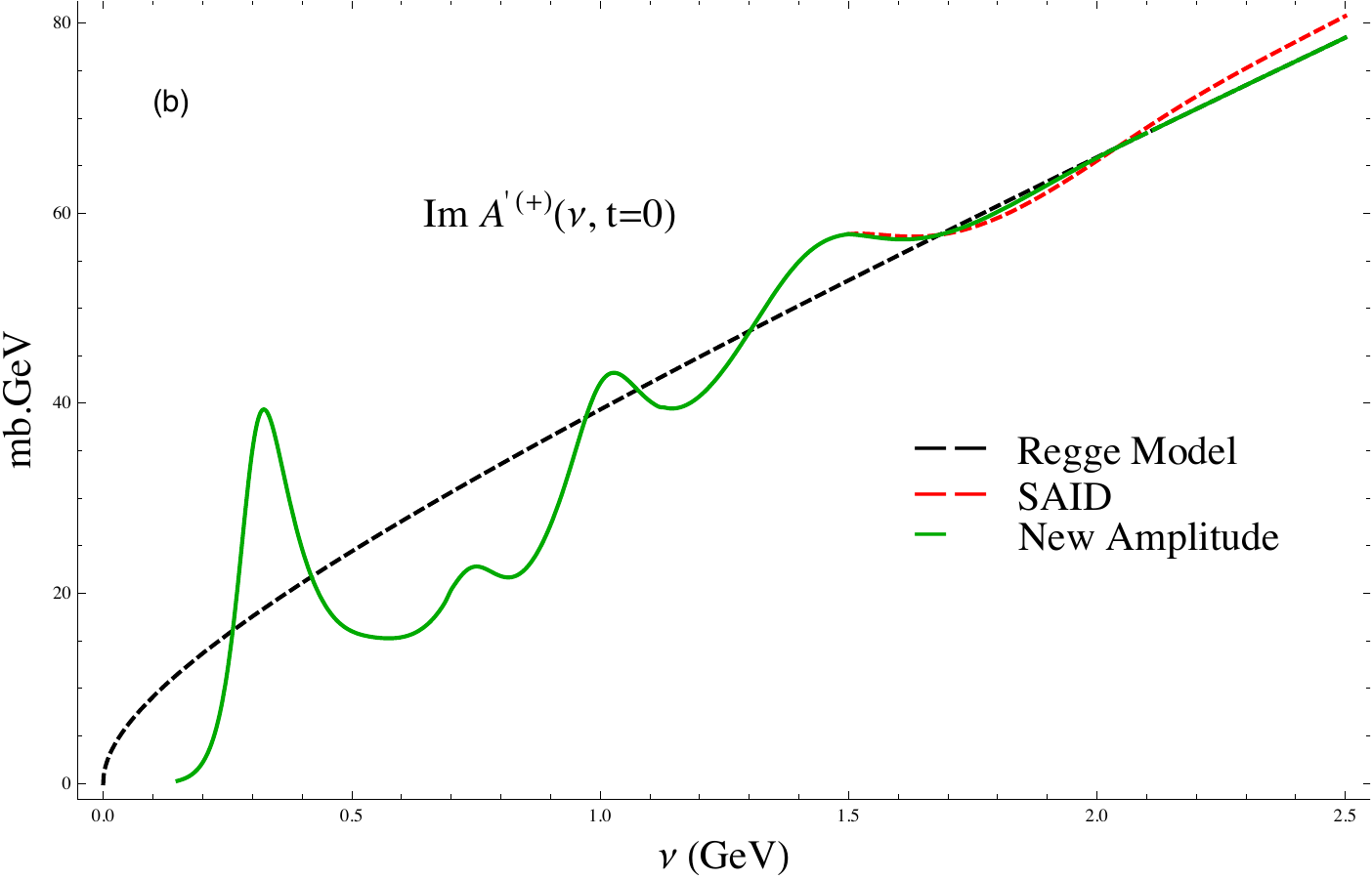}
 		\put (90,10) {\large$(b)$}
	\end{overpic}
	\begin{overpic}[width=0.49\textwidth]{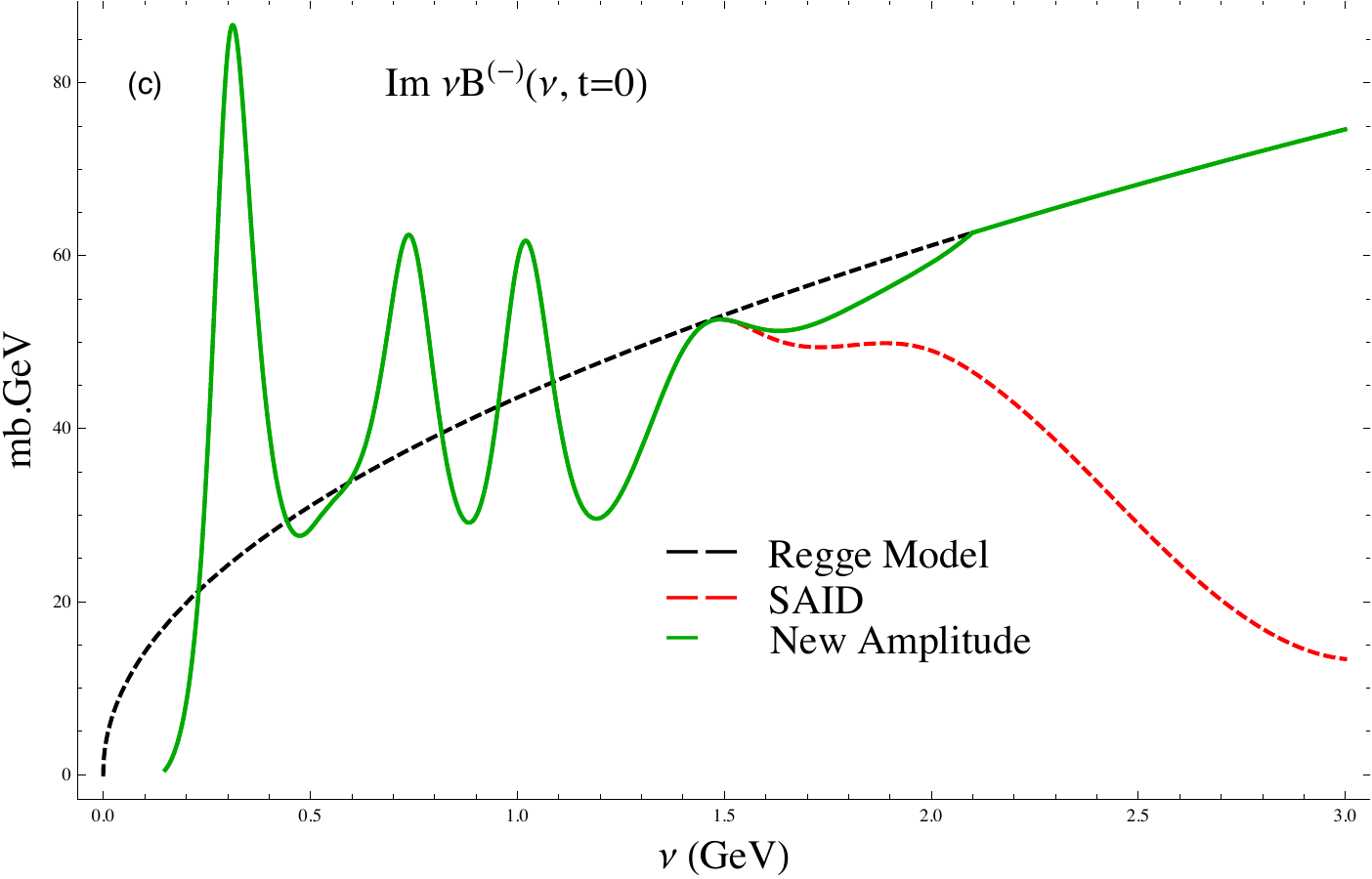}
 		\put (90,10) {\large$(c)$}
	\end{overpic}
	\begin{overpic}[width=0.49\textwidth]{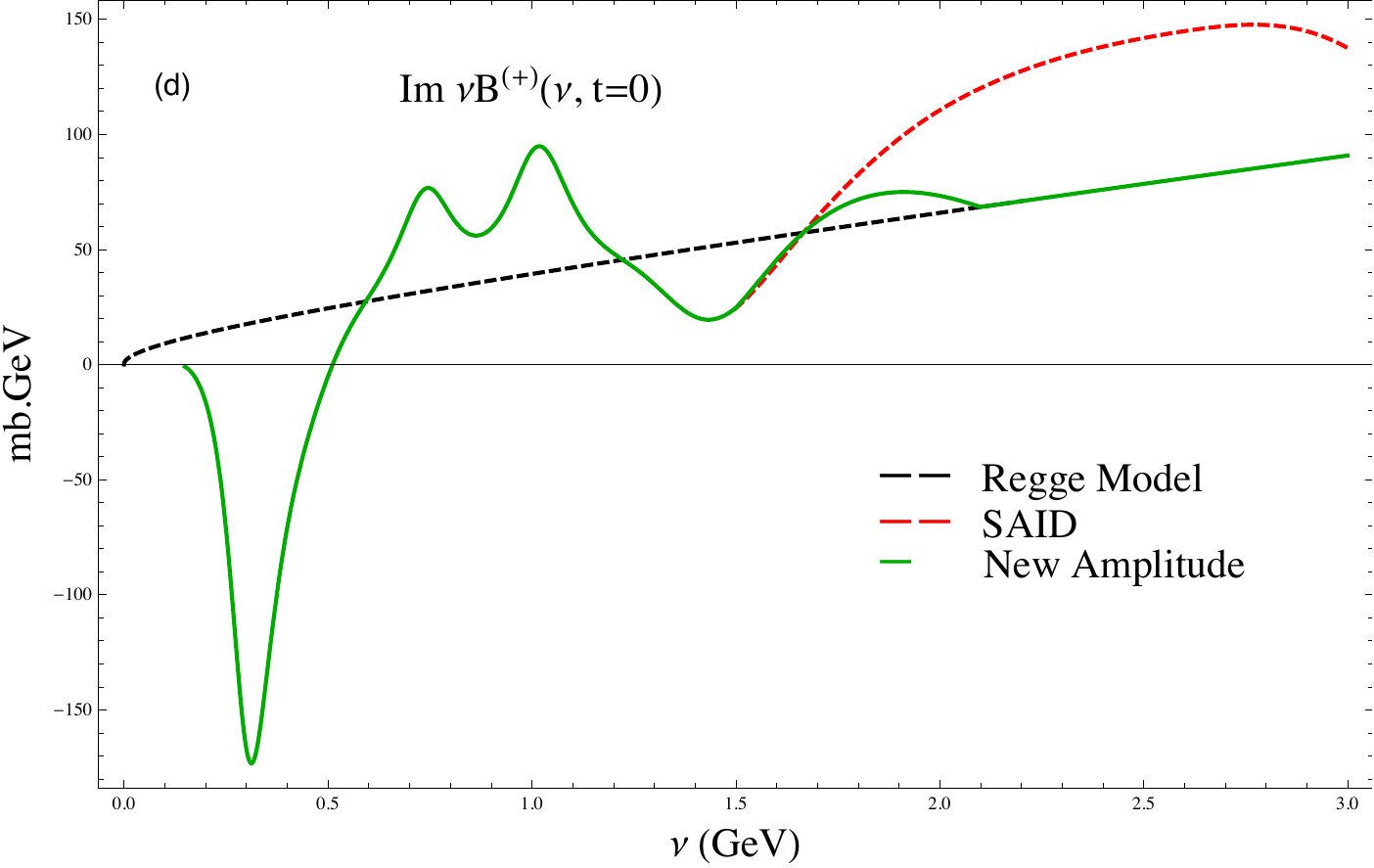}
 		\put (90,10) {\large$(d)$}
	\end{overpic}
\caption{\label{fig:NewIm}(color online) Matching low and high energy models in the intermediate region yields new amplitudes. Fig (a) amplitude $A'^{(-)}$. Fig (b) amplitude $\nu B^{(-)}$. Fig (c) amplitude $A'^{(+)}$. Fig (d) amplitude $\nu B^{(+)}$.}
\end{center}
\end{figure*}

In the following we illustrate this procedure using the SAID model at low energies $\nu < \nu_L \equiv 1.5\mbox{ GeV}$, and the Regge model described in Sec.~\ref{sec:high} at high energies, $\nu > \nu_H \equiv 2.1\mbox{ GeV}$. Between $\nu_L$ and $\nu_H$,  we use a linear interpolation. This simple method of connecting the two regions is enough for our purpose since the imaginary part of the amplitudes will be integrated in the dispersion relations. The resulting imaginary parts are shown Fig.~\ref{fig:NewIm}.

We reconstruct the real parts from the dispersion relations \eqref{eq:FTDR}. In the case of $A'^{(+)}$, because of the Pomeron,  the imaginary part grows like $\text{Im }A'^{(+)}\sim \nu^{1.075}$ at high energy.  The integrand in the dispersion relation therefore needs a subtraction. We choose to match the real part of the reconstructed $A'^{(+)}$ with the real part of the SAID amplitude at $s=1.5$ GeV$^2$. 

\begin{figure*}[htb]\begin{center} 
	\begin{overpic}[width=0.49\textwidth]{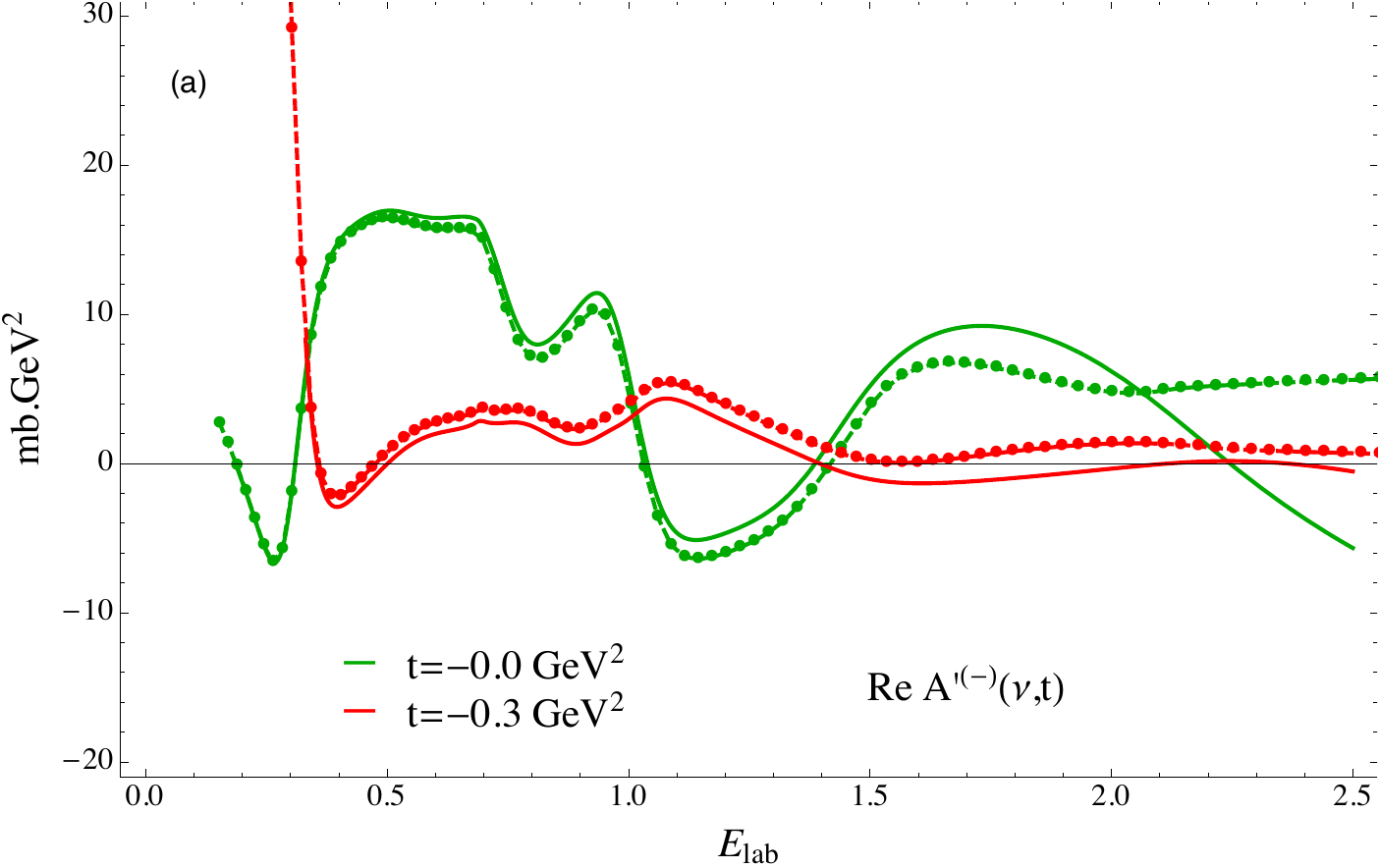}
 		\put (90,10) {\large$(a)$}
	\end{overpic}
	\begin{overpic}[width=0.49\textwidth]{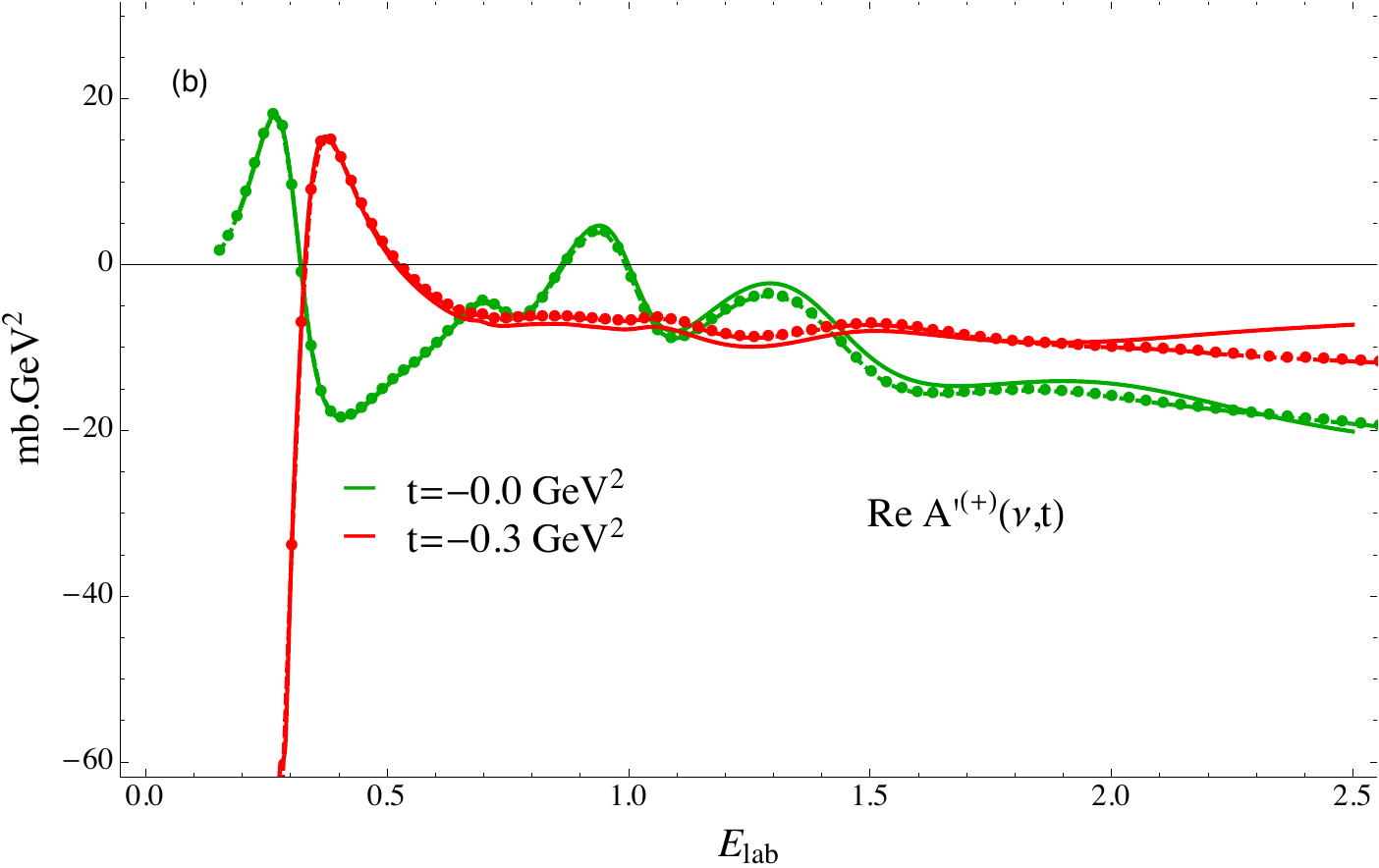}
 		\put (90,10) {\large$(b)$}
	\end{overpic}
	\begin{overpic}[width=0.49\textwidth]{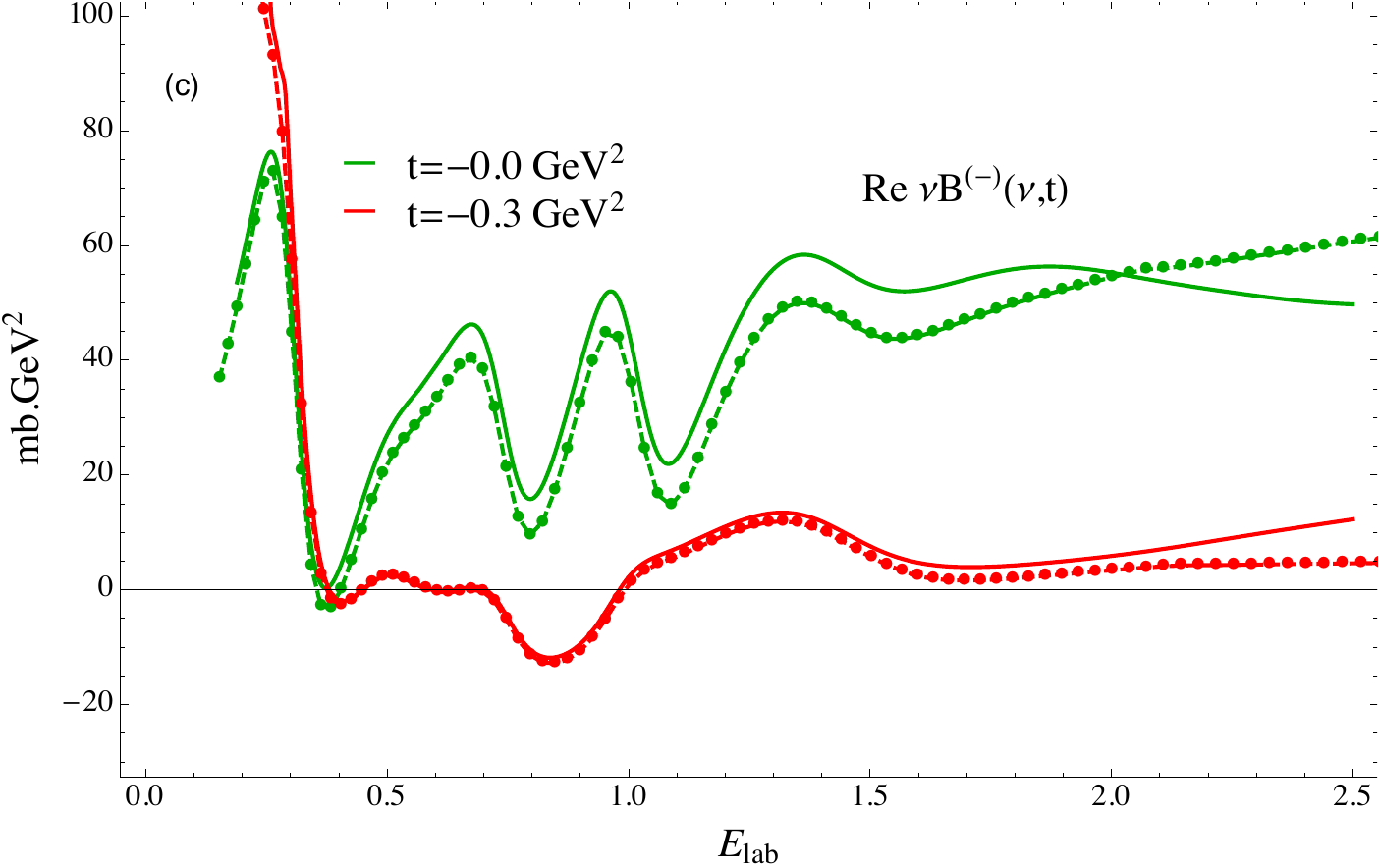}
 		\put (90,10) {\large$(c)$}
	\end{overpic}
	\begin{overpic}[width=0.49\textwidth]{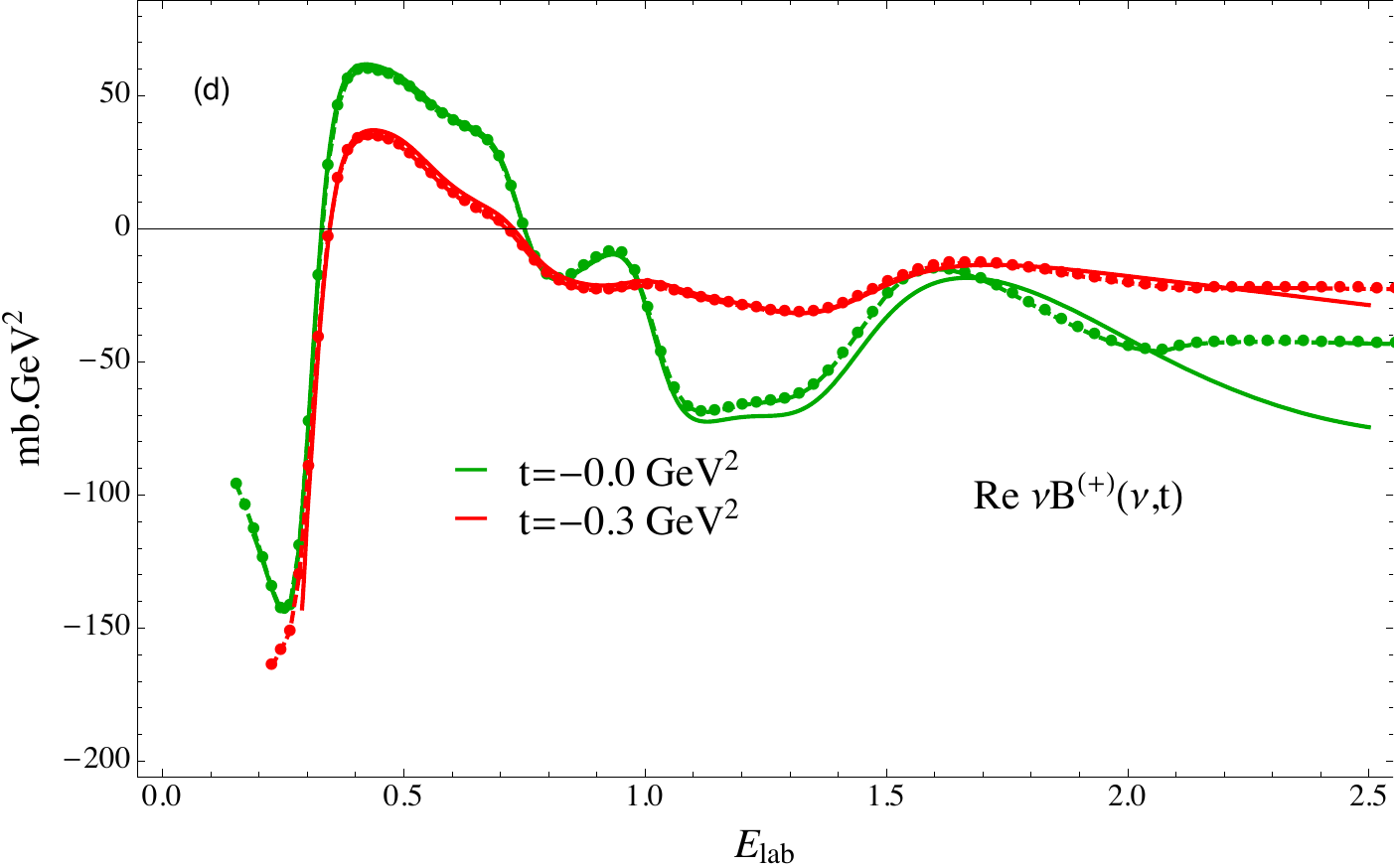}
 		\put (90,10) {\large$(d)$}
	\end{overpic}
  \caption{\label{fig:NewRe}(color online) The reconstructed real parts of amplitudes (dashed lines) is compared to SAID (solid lines) for $t=0$ (green) and $t=-0.3$ GeV$^2$ (red). Fig (a) amplitude $A'^{(-)}$. Fig (b) amplitude $\nu B^{(-)}$. Fig (c) amplitude $A'^{(+)}$. Fig (d) amplitude $\nu B^{(+)}$.}
\end{center}
\end{figure*}

In Fig.~\ref{fig:NewRe} we compare the real part of the new amplitudes and with those of SAID for $t=0$ and $t=-0.3$~(GeV$^2$). All four amplitudes globally agree. As expected the difference decreases as $t$ decreases because the FESR is better satisfied. 

In this study the high energy model was only constrained by the data and not the FESR. This can be improved by imposing both constraints simultaneously. 

Now than we have determined the real part of the new $\pi N$  amplitudes we could,
 in principle study the partial waves.  Inversion of the formulas in Eq.~ \eqref{eq:PW}, however, requires knowledge of the amplitudes in the whole domain of the scattering angle while in our study we focused on the $t-$channel Regge poles, which dominate the amplitudes in the forward direction, for $0<-t<1$~GeV$^2$ and $p_\text{lab}>3$ GeV.  
To extract the partial waves it would be necessary repeat the present analysis in the backward direction and include $u-$channel baryon exchanges which were studied, for example in  \cite{Barger:1967zzb, Lacombe:1973uu, Huang:2009pv}. This will be a subject of future analysis.

\section{Summary and future directions} \label{sec:summary}
Finite energy sum rules were  derived and applied for $\pi N$ charge exchange in the forward direction in the past \cite{Caltech, Logunov:1967dy, Igi:1967zza}. FESR's were later applied at finite-$t$ in charge exchange $\pi N$ to predict the $\rho$ exchange parameters \cite{Dolen:1967zz,Dolen:1967jr}.  More recently, the Bonn-Julich group compared their Regge amplitudes and SAID amplitude in the intermediate region \cite{Huang:2008nr,Huang:2009zzr}. The agreement is better for the spin flip amplitude compared to that for the non-flip.  The disagreement in the non flip amplitude may be related to the constraint on the residue being proportional to the trajectory. As we saw in Sec. \ref{sec:low} the zero in the non flip amplitudes is responsible for the cross over in $\pi^\pm p$ and appears at small $|t|$ and 
 not at the zero related to the wrong signature point $\alpha_\rho=0$.

In this work we investigated the possibility  of implementing the FESR constraints on a global fit to data. We first computed the finite energy sum rules from various solutions. They all displayed the same features. Guided by these results, we parametrized the high energy region with amplitudes involving the exchange of $t$-channel poles. The Pomeron and $f_2$ contributions to the $A'^{(+)}$ amplitude, with their magnitude constrained by the total cross section and their $t-$dependence constrained by this differential cross-section, satisfy the FESR very well. The FESR for the $B^{(+)}$ amplitude is not as well satisfied since we imposed the relation $\nu B^{(+)}=A'^{(+)}$ in the high energy region. The difference between the two side of the sum rules for $\nu B^{(+)}$ is however small. In addition we note that we compare the {\it r.h.s.} of the FESR with the {\it l.h.s} taken from SAID. When computed using other solutions, presented in Fig.~\ref{fig:FESRSolt}, the {\it r.h.s.} of the sum rule  lead to a similar agreement with our {\it l.h.s.}. The sum rule for the dominant isovector amplitude $\nu B^{(-)}$ is also very well satisfied. The largest relative deviations between the two sides of the sum rule are observed in the smallest amplitude $A'^{(-)}$. In particular the lowest moment of the left side of the sum rule for $A'^{(-)}$ displays a change of sign at a different $t$ with respect to its other moments. As we chose to reproduce the change of sign of the highest moments, the FESR for the lowest moment is not so well satisfied. In summary, an independent fit of the high energy data yield FESR's globally satisfied for the four amplitudes. There are nevertheless room for improvement. 

The transition region between resonances and Regge exchanges is found to be $E_\text{lab}\sim 1.6-2$ GeV in the forward direction. We joined the imaginary parts of the amplitudes in the two regions and defined new amplitudes in the whole energy range and for small angles. The real parts of these new amplitudes are reconstructed from the dispersion relation. The real parts compare well with the original SAID solution for small momentum transfers as shown in Fig.~\ref{fig:NewRe}. 

In practice, one would aim at implementing FESR's in a global amplitude fit. In such analysis the  low-energy region, parametrized through partial waves and the high-energy region, parametrized through Regge exchanges are fitted simultaneously, with FESR imposed as a constraint on fit parameters.  Imposition of FESR reduces model dependence of the low energy parametrization and might provide an additional check on systematic uncertainties  in extraction of baryon resonance parameters. 
  

Nowadays, SAID uses dispersion relations to constrain the real parts of the amplitudes~\cite{Arndt:1994bu,Arndt:1995bj,Arndt:2003if}. We expect that our Regge parametrization will help to implement, in a systematic way  finite energy sum rules in pion-nucleon scattering and reactions. With this aim,  all the material, including data and software are available in an interactive form online~\cite{website}.


\section{Acknowledgment} 
 We thanks R. Workman for many useful discussions about the SAID model. 
We thank M. Vanderhaegen for bringing the reference \cite{Huang:2009zzr} to our attention. 
 This material is based upon work supported in part by the U.S. Department of Energy, Office of Science, Office of Nuclear Physics under contract DE-AC05-06OR23177. This work was also supported in part by the U.S. Department of Energy under Grant No. DE-FG0287ER40365, National Science Foundation under Grant PHY-1415459.


\end{document}